\documentclass[apj]{emulateapj}

\shorttitle{Subdwarf B Models with MESA}
\shortauthors{Schindler et al.}

\begin{document}
 
 \title{Exploring stellar evolution models of sdB stars using MESA}
 
 \author{Jan-Torge Schindler, Elizabeth M. Green, and W. David Arnett}
 \affil{Steward Observatory, University of Arizona, 933 North Cherry Avenue, Tucson, AZ 85721, USA}

\begin{abstract}
Stellar evolution calculations have had great success reproducing the
observed atmospheric properties of different classes of stars.
Recent detections of g-mode pulsations in evolved He burning stars allow a rare comparison of their internal structure with stellar models.
Asteroseismology of subdwarf B stars suggests convective cores of $0.22-0.28\,M_\odot$, $\gtrsim 45\,\%$ of the total stellar mass. 
Previous studies found significantly smaller convective core masses ($\lesssim 0.19\,M_\odot$) at a comparable evolutionary stage.

We evolved stellar models with MESA (Modules for Experiments in Stellar Astrophysics)
to explore how well the interior structure inferred from asteroseismology can be reproduced by standard algorithms.
Our qualitative evolutionary paths, position in the $\log g-T_{\rm{eff}}$ diagram and model timescales are consistent with 
previous results. SdB masses from our full evolutionary sequences fall within the range of the empirical sdB mass 
distribution, but are nearly always lower than the median.

Using standard MLT with atomic diffusion we find convective core masses of $\sim 0.17-0.18 M_\odot$, averaged over the entire sdB lifetime.
 We can increase the convective core sizes to be 
 as large as those inferred from asteroseismology, but  only for extreme values of the overshoot parameter (overshoot gives
 numerically unstable and physically unrealistic behavior at the boundary).
High resolution three-dimensional (3D) simulations of turbulent convection in stars suggest that the Schwarzschild criterion for convective mixing sytematically {\em underestimates the actual extent of mixing because a boundary layer forms.} Accounting for this would decrease the errors in both sdB total and convective core masses.

\end{abstract}

\keywords{(stars: subdwarf) - stars: interiors - convection} 

\section{INTRODUCTION}


With the increasing improvement of stellar observations and the
introduction of new types of observations, it becomes possible to
constrain the processes involved and test the validity of the underlying
physics. Here we focus specifically on subdwarf B (sdB) stars which have 
been detected to have g-mode pulsations, 
and we use the Modules for Experiments in Stellar Astrophysics (MESA) code
\citep{Paxton2011, Paxton2013} for comparison with observational
constraints, with particular emphasis on mixing in the deep interior, a major unsolved 
problem in stellar evolution.

\subsection{Subdwarf B stars - General introduction and observational results}

Subdwarf B stars are hot ($T_{\rm{eff}} = 20,000 - 40,000\,\rm{K}$)
and compact ($\log g = 5.0 - 6.2$) stars that are found in all stellar
populations of our own Galaxy as well as in other old
galaxies. Located on the extreme horizontal branch (EHB), they are
understood to be helium burning objects with very thin hydrogen
envelopes ($ M_{\rm{H}}< 0.01 M_\odot$) \citep{Heber1986,
  Saffer1994}. While it is well known that they will directly evolve to
become white dwarfs once their central helium is exhausted, the
details and the relative importance of the various evolutionary
channels leading to the EHB are still poorly understood.  Numerous
single and binary star scenarios have been proposed \citep{Mengel1976,
Castellani1993, D'Cruz1996, Han2002, Han2003, MillerBertolami2008}.

Many sdB stars exhibit stellar pulsations, the shorter
($100-400\,\rm{s}$) pressure(p)-mode pulsations found in hotter V361
Hya stars \citep{Kilkenny1997} and the longer ($2000-14000\,\rm{s}$ or longer)
gravity(g)-mode pulsations in cooler V1093 Her stars
\citep{Green2003}, or both, in hybrid pulsators called DW Lyn stars
\citep{Schuh2006}.

Detailed work has recently been done to reproduce the observed pulsational
properties of sdB stars with stellar evolution models.
It has been shown that time-dependent diffusive processes, including mass loss,
gravitational settling, and radiative levitation 
\citep{ Chayer2004, Fontaine2006a, mrr07, mrr08, Hu2009, Hu2011}
are important to recover the iron-group opacity bump that excites the pulsations 
\citep{Charpinet1997} in these stars, as well as the correct position of the 
instability strip in the $\log g - T_{\rm{eff}}$ diagram 
\citep{ Fontaine2006b, Hu2008, Hu2009, Hu2010, Hu2011, Bloemen2014}.

Pulsational frequencies derived from light curves of p- and g-mode sdB
pulsators using ground-based as well as space-borne (i.e.\ CoRoT and
Kepler) instruments have been analyzed by asteroseismology.  Using the
forward method \citep[e.g.][]{VanGrootel2008, Charpinet2008}, the
parameters derived by asteroseismology, in particular total stellar
masses, surface gravities and effective temperatures, agree remarkably
well with measurements from other techniques such as light curve
modeling of eclipsing binary systems and spectroscopic analyses
\citep[Table~3]{VanGrootel2013, Green2011}.

Asteroseismology of g-mode pulsators also provides observational
benchmarks for the inner structure of those stars, e.g.\ the extent of
the inner convection zone and the nature of the abundance
gradient {\bf above} the boundary. 
\citet{VanGrootel2010a, VanGrootel2010} and
\citet{Charpinet2011} recently derived sdB He-CO convective core
masses for the first time from asteroseismology of three
g-mode pulsators: $M_{\rm{cc}} = 0.22\pm0.01\,M_\odot$,
$M_{\rm{cc}} = 0.28\pm0.01\,M_\odot$, and either $M_{\rm{cc}} =
0.274^{+0.008}_{-0.010}\,M_\odot$ or $M_{\rm{cc}} =
0.225^{+0.011}_{-0.016}\,M_\odot$.\footnote{The asteroseismological analysis
of \citet{Charpinet2011} identified two equally
probable solutions.}  All three stars
were determined to be significantly less than halfway 
through their He-burning 
lifetimes, having consumed only about 20\% to 40\% of the helium fuel in their 
cores.

Standard stellar evolution, without any additional mixing at the
boundary of the convective core, predicts a constant convective core mass
of $\sim 0.1\,M_\odot$ for these stars.  When some form of overshoot and 
semiconvection \citep{Sweigart1987}, partial mixing \citep{Dorman1993a} or 
atomic diffusion \citep{mrr07} is included\footnote{We note that atomic diffusion is physically well-defined, 
while overshoot and mixing are approximations to assumed fluid flow behavior, 
and are therefore likely to be more uncertain.}, 
the models produce 
growing convective cores, with masses up to $M_{\rm{cc}} \sim 0.17$ to $0.19\,M_\odot$
at an evolutionary stage comparable to the sdB stars whose convective cores were inferred
from asteroseismology.
Sweigart and Dorman achieved semiconvective or partially mixed cores
that extended up to $\sim 0.25\,M_\odot$ by the end of the He core burning lifetime, but
the observations indicate that such large cores are achieved much earlier. 

The forward modeling method implemented by both Van Grootel 
and Charpinet to analyse sdB asteroseismic data uses static stellar
models covering a much larger range of parameter space than
theoretical models that are constrained to lie along evolutionary tracks.
Since their results are independent of evolutionary calculations, this
allows us to test various physics options in a state-of-the-art
stellar evolution code (MESA).

For this paper, we calculated a series of sdB stellar evolution
models to compare with observational results derived from both spectroscopy and
asteroseismology. In \S\,\ref{method}, we explain the method and 
assumptions in our stellar evolution calculations.  In \S\,\ref{models_section}, 
we present our results, discuss the agreement with atmospheric parameters derived 
from spectroscopic observations and compare with previous sdB models.
\S\,\ref{conv_core_section} focuses on the convective cores of sdB stars and
the effects of different input physics on the extent of the inner convection zone
and the convective core evolution.
We summarize our findings in \S\,\ref{conclusion}.

\subsection{Mixing processes}

Turbulent convection is an essential process of energy transport in
stars. Macroscopic mass elements start to rise (or sink) in
dynamically unstable regions, delivering their excess (or deficit)
of heat to cooler (or hotter) layers and thus transport energy and
material throughout the star. This is a non-linear process governed by
the Navier-Stokes equation, and occurs on the dynamical
timescale. 

\subsubsection{Canonical mixing}
MESA treats the mixing of convective
elements as a ``diffusive'' process, as do most stellar evolution codes. 
This ``diffusion'' operator is chosen for mathematical convenience 
\citep{Eggleton1972}, 
based on parameters estimated from
the mixing length theory \citep{Bohm-Vitense1958} 
as formulated by \cite{cox1968principles}.  

To estimate the extent of dynamically unstable regions, two criteria
of linear stability are implemented in MESA. The standard Schwarzschild
criterion for instability,
\begin{equation}
 \nabla_{\rm{rad}} > \nabla_{\rm{ad}} \label{schw}
\end{equation}
and the Ledoux criterion, taking radial composition gradients $\nabla_\mu$ into account,
\begin{equation}
 \nabla_{\rm{rad}} > \nabla_{\rm{ad}} + \frac{\phi}{\delta} \nabla_\mu \ ,
\end{equation}
where 
\begin{equation}
\phi := \left. \left( \frac{\partial \ln \rho}{\partial \ln \mu} \right) \right|_{P, \mu}  \delta := -\left. \left( \frac{\partial \ln \rho}{\partial \ln T} \right)\right|_{P, T} \ .
\end{equation}
Historically, convective mixing was supposed to occur in the region in which the Schwarzschild condition (Eq.~\ref{schw}) was violated; we will call this {\it canonical mixing}. 
In addition to this canonical mixing, it has proven desirable to consider additional mixing processes (some of which are not truly different, but overlooked in the stellar approximation to turbulent hydrodynamics).

\subsubsection{Overshoot}

The term {\it overshoot} refers to the transport of energy and material
across the boundary from the dynamically unstable region into the
dynamically stable region. The additional mixing is calculated using
the diffusion coefficient from the previous MLT calculations near the
boundary layer and extrapolates it into the
radiative region with an exponential decay, 
following \cite{Herwig2000}; 
this procedure is based on simulations of fluids in shallow convection zones, of homogeneous
composition \citep{fls96}. It has no compositional dependence. 
The additional term is referred to as the overshoot mixing diffusion coefficient,
\begin{equation}
 D_{\rm{OV}} = D_{\rm{conv,0}} \exp \left( - \frac{2 \Delta r }{f_{\rm{ov}} \lambda_{P,\rm{0}}} \right) 
\end{equation}
where $D_{\rm{conv,0}}$ is the previously calculated diffusion
coefficient at a user defined location close to the Schwarzschild
boundary, $\lambda_{P,\rm{0}}$ is the local pressure scale height,  
and $\Delta r$ is the distance of overshoot into the radiative layer. 
The local pressure scale height is the exponential attenuation length for g-mode
waves \citep{LL-FluidMechanics}, and it is assumed that overshoot falls off
as wave energy.
The free parameter
$f_{\rm{ov}}$ sets the extent of the overshooting region and needs to
be adjusted by the user depending on the problem.
 
In \cite{Paxton2013} an $f_{\rm{ov}}$-parameter of $f_{\rm{ov}} =$ 0.004
to 0.015 was used to calculate models of a non-rotating $1.5M_\odot$
star. \cite{Herwig2000} used an overshoot parameter of $f_{\rm{ov}} =
0.016$ for his study of $3M_\odot$ and $4M_\odot$ asymptotic giant branch (AGB) stars.

\subsection{Atomic diffusion and radiative levitation}

Georges Michaud led in the application of true {\it diffusion} processes and radiative 
levitation to stellar evolution models \citep{michaud70,michaud91}.
These processes have been applied to horizontal branch (HB) and sdB stars by
 \citet{ Michaud1985, Bergeron1988, Fontaine1997, Chayer2004, Fontaine2003, Fontaine2006a, 
Fontaine2006b, mrr07, mrr08, Hu2008, Hu2009, Hu2010, Hu2011, mrr11, Bloemen2014}.

\citet {Hu2011} defined atomic diffusion to include gravitational settling, thermal diffusion,
concentration diffusion, and radiative levitation.  Atomic diffusion in MESA
includes all of these processes except radiative leviation, which is a separate option.
Gravitational settling and radiative levitation
are particularly important to recover the iron-group opacity bump 
that excites the pulsations in these stars \citep{Charpinet1997}, the 
correct position of the instability strip in the $\log g - T_{\rm{eff}}$ diagram \citep{Bloemen2014}, 
and help in understanding their observed atmospheric abundances \citep{mrr11}. 
We will further examine the effects of atomic diffusion on the deep interior 
in \S\,\ref{conv_core_section}.

\section{STELLAR EVOLUTION CALCULATIONS} \label{method}

\subsection{Subdwarf B modeling using MESA}

The stellar evolution calculations were done with version 7184
of MESA \citep{Paxton2011, Paxton2013} in order to  
extend the results of \citet{Ostensen2012}. MESA
offers a variety of up-to-date physics modules,
including several MLT convection options, and is capable of
evolving stars through the He-flash, a crucial part of the
evolutionary path to sdB stars. The latter is modeled as a
quasi-static process, with MLT mixing, as a substitute for the full dynamical process.

Because our main goal was to devise a simple model that allowed us to focus 
on the inner convection zone of the star, we
used standard values and descriptions for the input physics, varying 
the defaults for consistency with the observational data, 
as described below. 

The physics options adopted for our standard models are summarized
in Table~\ref{simulationphysics}.  They
include an atmospheric boundary condition of $\tau = 2/3$ and a mixing 
length parameter of $\alpha_{\rm{MLT}}=2$.
We further selected a nuclear network designed to include all reactions for
hydrogen and helium burning.  Representative abundances
of $Z= 0.02$, $Y = 0.28$ and $X = 0.70$ were adopted, because most well-studied 
(i.e.\ nearby) sdB stars belong to the field population of the galactic disk 
\citep{1991PhDT........10S, Saffer1994}, and the few that are found in
old open clusters (where their progenitors' abundances can be measured) 
appear to be preferentially metal-rich (see \S\,\ref{obs constraints}).
We used the Reimers wind scheme \citep{Reimers1975} with
$\eta_{\rm{Reimers}} = 0.5$ on the red giant branch (RGB). For the
post-EHB phase we used the Bl\"ocker wind scheme \citep{Blocker1995}
with $\eta_{\rm{Bl\ddot{o}cker}} = 0.5$.  

MESA offers the opacity tables of \citet{Iglesias1993, Iglesias1996} and 
includes their OPAL type I opacity tables with fixed metal distributions 
as the default option.
However, the interior abundances of carbon and oxygen in sdB stars 
change enough to modify the opacity, not only during He burning on the EHB, but also
during their prior evolution, in particular during the He flash and the
subsequent transition to the ZAEHB.  
It is therefore necessary to use the OPAL type II opacity tables
which allow for time dependent variation of the C and O abundances.

MESA includes electron conduction opacities \citep{Cassisi2007} as the default case. 
Electron conduction becomes the dominant energy transport mechanism in significantly degenerate stellar cores and is thus required for our study. 

Mixing due to atomic diffusion has been traditionally assumed to be small 
during stellar evolution, but it is necessary to obtain the correct 
structure to excite pulsations in sdB stars and reproduce the observed
instability strips \citep{Fontaine2003, Fontaine2006a, 
Fontaine2006b, Hu2009, Hu2011, Bloemen2014}. 
As chemical diffusion smoothes out abundance gradients at the boundary of 
the convective core, this diffusion, combined with time dependent opacities, 
allows the convective core to grow
during the sdB evolution. This was first noted by \citet{mrr07} in the context 
of somewhat more massive, but essentially similar, HB stars.

Hence we included type II opacities as well as atomic diffusion processes 
in all of our stellar evolution calculations, unless otherwise specified.  
We deliberately did not include radiative levitation in our set of standard 
parameters since it is computationally very expensive and 
has little to no effect on the interior mixing regions  of the sdB models
(see \S\,\ref{conv_core_section}).

MESA calculates the enthalpy flux from standard mixing length theory
\citep{cox1968principles}, using either the Schwarzschild or Ledoux
stability criterion. With the latter, semiconvection may be included
\citep{Langer1983}. Our initial investigation of the effects of
semiconvection on the convective cores of sdB stars showed that
semiconvection, as implemented in MESA, has little effect on the convective core sizes.
The Ledoux criterion with maximal semiconvection gives convective core sizes similar to those obtained with the Schwarzschild criterion. 
Convective overshoot is implemented according to 
\citet{Herwig2000}. In \S\,\ref{conv_core_section}, we investigate 
the effect of varying amounts of overshoot on the convective cores.

\begin{table}[ht]

\begin{center}
\caption{Specified physics for standard set of stellar evolution models}
\label{simulationphysics}
\begin{tabular}{cc}
\tableline
\tableline
\\
Opacity & OPAL type II \\ 
 & electron conduction \\
Nuclear network & \texttt{pp\_cno\_extras\_o18\_ne22.net} \\
Metallicity & Z = 0.02 \\
Composition & $Y = 0.30$, $X = 0.68$ \\ 
MLT  & $\alpha_{\rm{MLT}}$=2 \\
RGB wind scheme & $\eta_{\rm{Reimers}} = 0.5$ \\
AGB wind scheme & $\eta_{\rm{Bl\ddot{o}cker}} = 0.5$ \\
Convection criterion & Schwarzschild \\
Diffusion options & Atomic diffusion \\
\\
\tableline
\end{tabular}
\end{center}

\end{table}

\subsection{Our method}

We modeled sdB stars that have evolved from solar type stars in
binary systems through the common envelope (CE) or the Roche lobe
overflow (RLOF) channel \citep{Han2002}. These apparently very common
scenarios are believed to occur in binary systems where the companion
strips away the hydrogen envelope of the expanding progenitor star as
the latter evolves toward the tip of the RGB.  When most of the red
giant envelope is removed, hydrogen shell burning is quenched, the
He-core stops growing, and the star begins to contract away from the
RGB.  If the helium core is sufficiently massive for the contraction
to trigger helium ignition, the star will evolve onto the EHB and
begin burning He in its core as an sdB star.

For simplicity, we simulated the effect of binary mass
stripping by removing the envelopes of non-rotating single stars.
Both spectroscopy \citep[e.g.\,][]{Heber2000} and asteroseismology
\citep[e.g.\,][]{Randall2007, Baran2012} indicate that sdB stars are
generally slow rotators, except those that have been spun up to some
degree by a binary companion.  Our assumption that rotation does not
play an important role even in the interiors of sdB stars is based on
available asteroseismic evidence, which suggests rigid rotation in
both binary and single sdB stars for the few cases so far in which
the interior rotation could be constrained
\citep{Charpinet2008, VanGrootel2008, Pablo2012, Charpinet2011a}.

Our first step was to create pre-main-sequence (PMS) models, which
are specified by their initial mass $M_{\rm{ini}}$, a uniform
composition, a luminosity and a central temperature ($T_c = 9 \times
10^5~\rm{K}$ by default). Once the PMS routine found the central
density $\rho_c$ that gives the model the desired mass, we evolved the
star up to the point of the He-flash. Just before the flash occurred,
we saved the structure as our sdB progenitor model.  
Our procedure provides an {\it upper} limit to the He core mass
of the resulting sdB star, since the progenitor could have been
stripped prior to the He flash when the He core mass was up to $\sim
0.02\, M_\odot$ smaller and it would still have evolved onto the EHB
\citep{Castellani1993, D'Cruz1996}.

For the next step, we stripped away mass from the sdB progenitor model
beginning with the outermost cell using the \texttt{relax\_mass}
option of MESA. This option ensures that the mass of the star is
adjusted to the specified value of \texttt{new\_mass}. The mass loss
occurs in a series of small episodes until the requested new mass is
reached. Then the code begins the actual stellar evolution toward the
zero age EHB (ZAEHB).  Mass loss continues to occur between the RGB
tip and the ZAEHB.  Using this method, we reduced the
hydrogen envelope down to typical values for sdB stars in the range of
$M_{\rm{H}} = 10^{-4} {\rm\ to\ } 10^{-3}\,M_\odot$.

This simple mass loss procedure differs from the real dynamics 
in that it is modeled as a quasi-static process. The full 3D
well-resolved hydrodynamics for a binary star over evolutionary
timescales is still beyond the capability of modern computation.
Therefore, following \citet{Dorman1993, Hu2009, Hu2011, 
Ostensen2012, Bloemen2014}, we adopted a more tractable approach. 

The evolutionary tracks of sdB stars are primarily influenced by two
factors, the mass of the He core after the envelope has been stripped,
and the amount of remaining hydrogen envelope.  As seen in previous
studies \citep{Dorman1993, Han2002}, the initial mass of the
progenitor star, $M_{\rm{ini}}$, determines the mass of the He core at
the He flash and therefore the approximate total mass of our sdB
stars, subject to a very small dependence on composition.  For ages
appropriate for old disk stars ($\lesssim 10$ Gyr), the more massive
the progenitor star, the earlier the star begins helium fusion and
thus the less massive the He core. 
For a given
He core mass, the tiny envelope mass is decided by the parameter
$M_{\rm{new}}$, the new stellar mass after stripping off the envelope.
Qualitatively speaking, the more massive the sdB envelope, the lower
the effective temperature and the surface gravity. In our
survey set of calculations, we varied the initial mass, $M_{\rm{ini}}$,
between $1.0{\rm \ and\ }2.5\ M_\odot$ and the new mass parameter,
$M_{\rm{new}}$, between $0.472{\rm \ and\ }0.490\ M_\odot$.

\section{SDB MODELS WITH CANONICAL MIXING AND ATOMIC DIFFUSION} \label{models_section}

In this section we compare the results of our sdB modeling with
observational data and with other stellar evolution calculations.  We
created a range of sdB models using standard MLT (i.e.\ 
excluding overshoot and semiconvection) and the Schwarzschild
criterion to explore the effects of varying the initial mass,
$M_{\rm{ini}}$, and stripped mass, $M_{\rm{new}}$, on our sdB models.

We show that our models are quite consistent with 
previous calculations of sdB stars.

The distances to field sdB stars are only approximate, so
luminosities cannot be accurately calculated. However, since effective
temperatures and surface gravities can be inferred from a variety of
methods and the masses of most sdB stars are very similar, they are
typically plotted in the $\log g - T_{\rm{eff}}$ plane instead 
of an HR diagram.

\begin{figure}[ht]
\begin{center}
 \includegraphics[width= 0.47\textwidth]{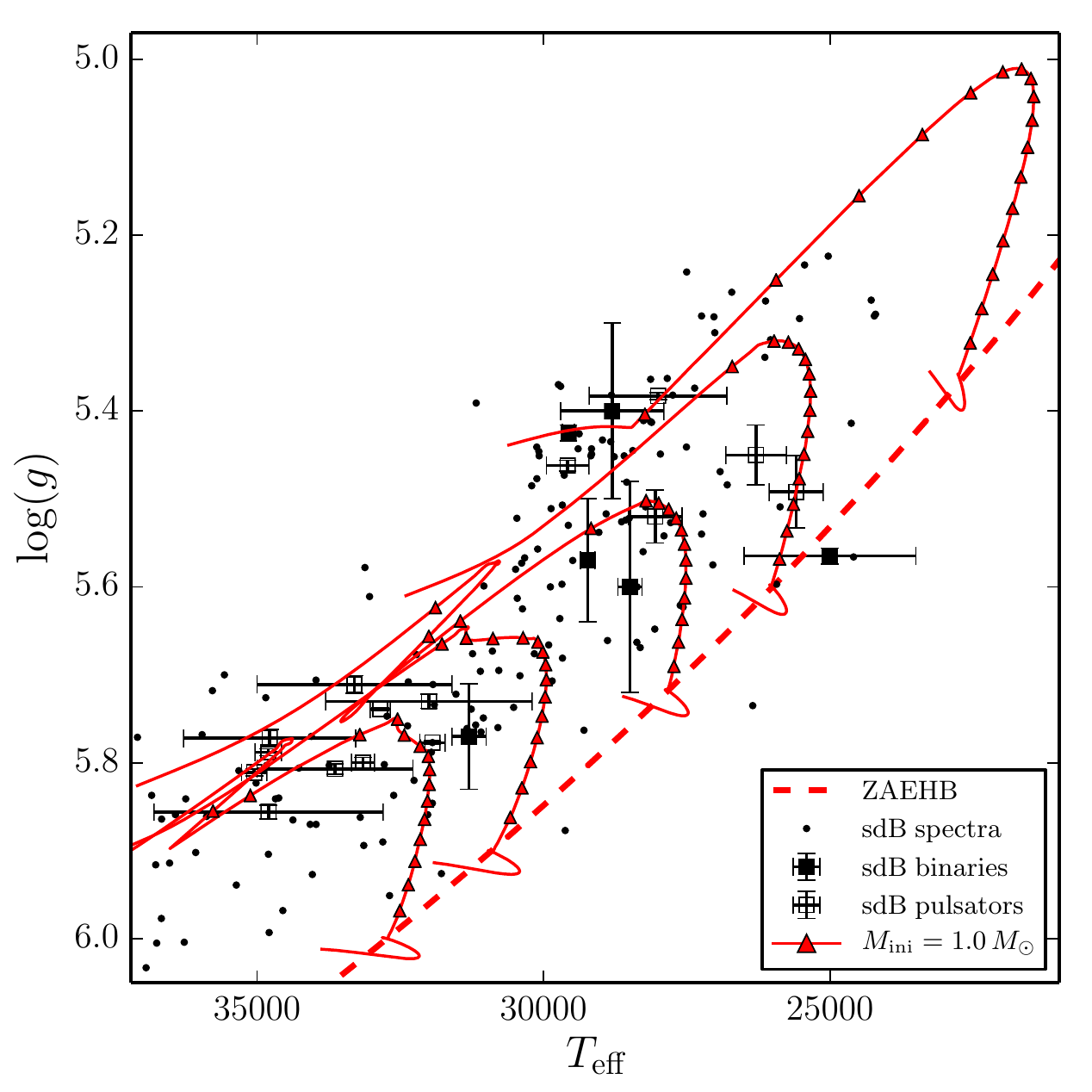} 
 \caption{Evolutionary tracks of sdB stars in the $\log g -
   T_{\rm{eff}}$ diagram. The solid black curves (red in the online
   journal) show the evolutionary tracks of our models with
   $M_{\rm{ini}} = 1.0\,M_\odot$ and $M_{\rm{sdB}}= 0.4652, 0.4654,
   0.4658, 0.4669, 0.4705\,M_\odot$ from bottom to top. The black
   (red) triangles intersect the lines in intervals of $10^7$ years,
   depicting the region where most sdB stars should be found. The
     small loop at the bottom of each track shows the final approach
     to the ZAEHB. The dashed lines show the ZAEHB's for our
   evolutionary models of $M_{\rm{ini}} = 1.0\,M_\odot$.  The
   spectroscopic data points (small black dots) for sdB stars 
   \citep{Green2008}, agree very well with the open and filled
     squares with error bars derived from eclipsing binary and
   asteroseismology analyses, respectively \citep{Fontaine2012}.}
 \label{sdBtracks1}
\end{center}
\end{figure}

Figure~\ref{sdBtracks1} shows the ZAEHB (lower dashed line) and
evolutionary tracks (solid lines) for our set of sdB models with 
$M_{\rm{ini}} = 1.0$ and a range of $M_{\rm{new}}$.
The tracks cover only the period of helium core
burning, when the star is characterized as an sdB star.
For comparison, we plot observed spectroscopic
data derived with the use of NLTE model atmospheres \citep{Green2008}, 
along with results from binary modeling and asteroseismology
\citep{Fontaine2012}.  
(Although Fontaine listed spectroscopic values for the effective
temperatures, because they are poorly constrained by asteroseismology,
we show the asteroseismic values for $T_{\rm{eff}}$
whenever they were available.)

Our models reproduce the characteristic hook-shaped sdB evolutionary
tracks, corresponding to stable helium core burning, as well as the
shape of the ZAEHB.  The tracks overlap the distribution of the
observational data points fairly well, except for a small overall
shift toward lower $T_{\rm{eff}}$ and/or higher $\log g$.

\begin{figure}[ht]
 \centering
 \includegraphics[width = 0.47\textwidth]{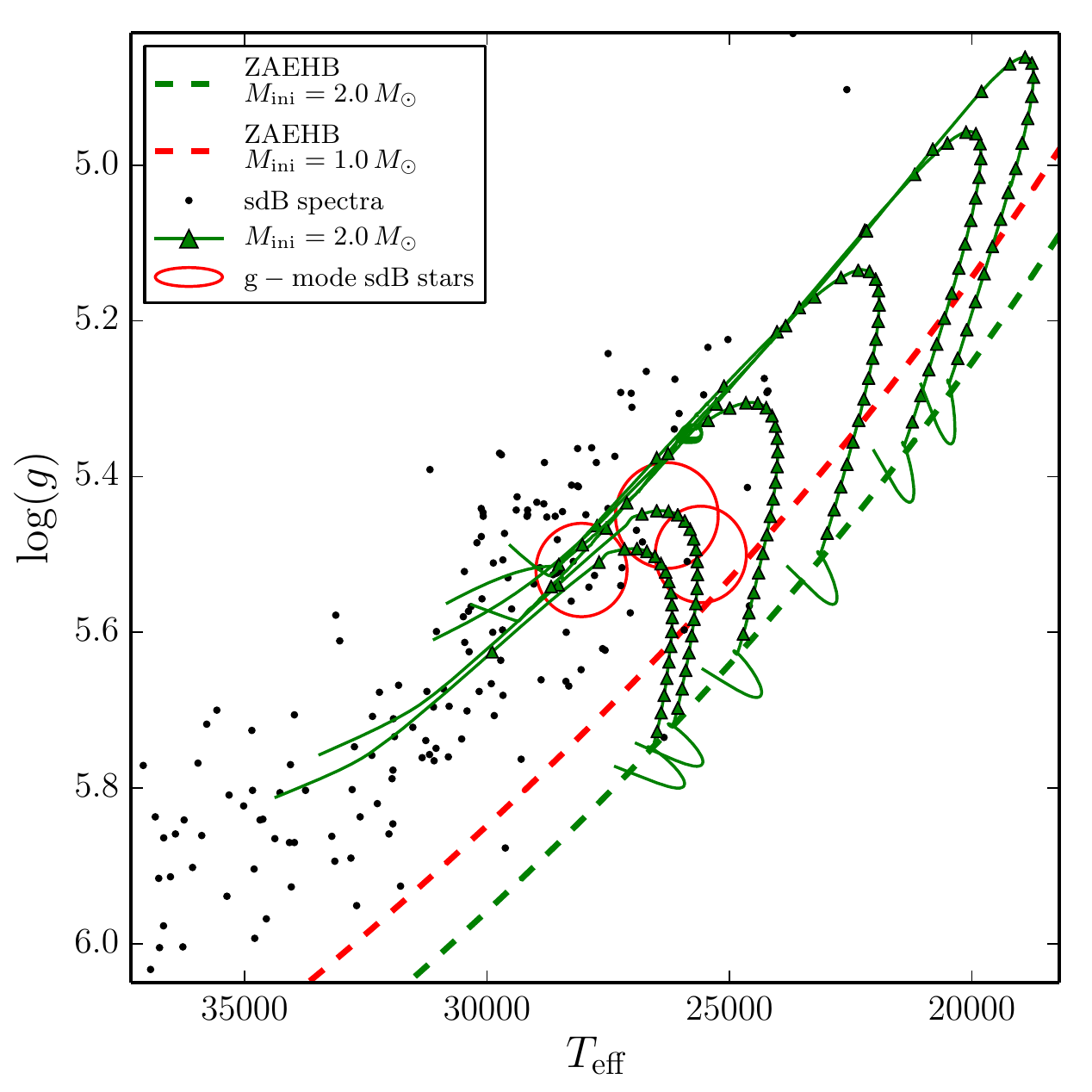} 
 \caption{Evolutionary tracks of sdB stars in the $\log g -
   T_{\rm{eff}}$ diagram. The solid black curves (green in the online
   journal) show the evolutionary tracks for sdB stars with
   $M_{\rm{ini}} = 2.0\,M_\odot$ and $M_{\rm{sdB}}= 0.4357, 0.4363,
   0.4373, 0.4389, 0.4415, 0.4433\,M_\odot$. The dashed lines show the
   ZAEHB for $M_{\rm{ini}} = 1.0\,M_\odot$ (upper/red) and
   $2.0\,M_\odot$ (lower/green).  The smaller He core masses developed
   by the more massive progenitors shift the ZAEHB for the
   $M_{\rm{ini}} = 2.0\,M_\odot$ models toward lower
   $T_{\rm{eff}}$/higher $\log g$ relative to the ZAEHB for
   $M_{\rm{ini}} = 1.0\,M_\odot$.  The observed spectroscopic data
   points (black dots) are the same as in Fig.\,\ref{sdBtracks1}.  The
   large open ellipses (black/red) correspond to the three g-mode sdB stars
   analyzed with asteroseismology, whose total masses are 0.496, 0.471
   and $0.463/0.452\,M_\odot$ (from left to right).}
 \label{sdBtracks2}
\end{figure}

Figure\,\ref{sdBtracks2} shows the ZAEHB (lower dashed line) and the
corresponding tracks (solid lines) for a different sequence of sdB
models evolved from an initial mass of $2.0\,M_\odot$. The resulting
He core masses, and therefore the total sdB masses, are about
$0.03\,M_\odot$ smaller than in the case of the $1.0\,M_\odot$
progenitor.  The ZAEHB for the $1.0\,M_\odot$ progenitor sequence of
Figure\,\ref{sdBtracks1} is shown (upper dashed line) for comparison.

It is also apparent from Figure\,\ref{sdBtracks2} that the three g-mode sdB stars
whose convective core masses were derived from asteroseismology
(plotted as open ellipses) have not evolved very far from the
ZAEHB's appropriate for their masses.

Details for the sequences of evolutionary tracks in
Figures\,\ref{sdBtracks1} and \ref{sdBtracks2} are given in
Table~\ref{1_2Msun_table}. The average sdB mass, the
total mass of hydrogen (essentially all in the envelope), 
and the extent of the inner convection zone of the sdB star are 
averaged over the time of helium core burning (the sdB lifetime).
We also provided the ZAEHB core and envelope masses for all stellar models. 

In agreement with previous HB/EHB stellar models, a smaller He
core mass with the same set of envelope masses shifts the ZAEHB in the direction
of lower effective temperatures and slightly higher surface gravities,
whereas a different envelope mass for the same core mass slides the
starting point of the sdB track up or down along the ZAEHB. 
Therefore, mass loss between the He flash and the ZAEHB results in 
hotter, higher-gravity sdB tracks, but the position of the ZAEHB does not change.

\begin{table*}[]
\begin{center}
 \caption{Properties for sdB tracks in Fig.\,\ref{sdBtracks1} with $M_{\rm{ini}} = 1.0$  (upper section) and in Fig.\,\ref{sdBtracks2} with $M_{\rm{ini}} = 2.0$ (lower section) }

\begin{tabular}{ccccccc}
\tableline
\tableline
\\
                & ZAEHB         & ZAEHB &          &               &                & Average\\
New mass	& core    & envelope  & Average  & Average sdB   & SdB lifetime	& convective\\
                &  mass& mass & sdB mass & hydrogen mass &                & core mass\\
 $M_{\rm{new}} [M_\odot]$ & $ [M_\odot]$ & $ [10^{-3}\,M_\odot]$ &$ M_{\rm{sdB}} [M_\odot]$ & $M_{\rm{H1}} [10^{-3}\,M_\odot]$ & $[10^6\,\rm{yrs}]$ & $M_{cc} [M_\odot]$ \\
 \\
\tableline
\\
0.4720	&	0.4596	&$5.722^\dag$  &	0.4652	&	0.23	&	160.7	&	0.181	\\
0.4730	&	0.4630	&$2.451^\dag$  &	0.4654	&	0.25	&	148.9	&	0.179	\\
0.4740	&	0.4650	&0.676 &	0.4656	&	0.25	&	160.9	&	0.179	\\
0.4750	&	0.4651	&0.758 &	0.4658	&	0.30	&	162.2	&	0.179	\\
0.4760	&	0.4652	&0.903 &	0.4660	&	0.41	&	142.6	&	0.173	\\
0.4800	&	0.4652	&1.813 &	0.4669	&	0.95	&	146.8	&	0.172	\\
0.4850	&	0.4650	&5.780 &	0.4705	&	3.46	&	171.3	&	0.173	\\
0.4900	&	0.4648	&10.365&	0.4749	&	6.27	&	174.4	&	0.170	\\
\\
\tableline
\\
0.4360	&	0.4350	&0.772&	0.4357	&	0.34	&	193.3	&	0.161	\\
0.4380	&	0.4354	&0.978&	0.4363	&	0.46	&	194.3	&	0.161	\\
0.4420	&	0.4356	&1.809&	0.4373	&	0.99	&	201.3	&	0.161	\\
0.4450	&	0.4355	&3.595&	0.4389	&	2.11	&	220.5	&	0.163	\\
0.4480	&	0.4354	&6.328&	0.4415	&	3.86	&	227.0	&	0.161	\\
0.4500	&	0.4353	&8.208&	0.4433	&	5.03	&	226.3	&	0.159	\\
\\
\tableline 
\\
\multicolumn{7}{c}{ 
\begin{minipage}{1.4\columnwidth}
\dag During the He flash hydrogen is mixed down in these two models, which gives larger envelope masses due to our envelope definition ($X_{\rm{H}}>0.1$). The preflash envelope masses are $0.468\cdot10^{-3}\,M_\odot$ for the first and $0.505\cdot10^{-3}\,M_\odot$ for the second model.
\end{minipage}
}
\end{tabular}
\label{1_2Msun_table}
\end{center}
\end{table*}

We ran additional sets of simulations for sdB stars with $M_{\rm{ini}}
= 2.3$ and $2.5\,M_\odot$ progenitors (not shown). Their helium core
masses are even less massive and thus the ZAEHB and sdB evolutionary
tracks are further shifted to even lower temperatures/higher gravities. In
these models the helium core mass is so low that the stars start
helium fusion before the material in the helium core becomes
degenerate.

\subsection{Comparison with other evolutionary models and observations} \label{model_comparison_section}

We compared our results with older evolutionary tracks computed
by Dorman for \citet{Charpinet2000, Charpinet2002c}, and with recent
work by \citet{Ostensen2012}, \citet{Bloemen2014}, and Fontaine (priv.\ comm.).

Dorman's primary set of models \citep{Charpinet2002c} adopted a He core mass of $0.4758\,M_\odot$.
We could not achieve that high an evolutionary
core mass with our standard set of parameters, 
for the same solar composition.  Therefore, we used MESA to construct a new ZAEHB,
starting from a He main-sequence star with a $0.4758\,M_\odot$ core
mass and using OPAL type II opacities.  Figure\,\ref{dormancomparison} shows that 
this ZA\-EHB is completely consistent with the starting points of Dorman's sdB tracks. 
Our model sdB tracks from Figure\,\ref{sdBtracks1} ($M_{\rm{ini}} = 1.0\,M_\odot$) also agree well with Dorman's,
spanning essentially the same range in $\log g - T_{\rm{eff}}$ space, although
ours are slightly shifted to lower temperatures and gravities due to  
$\sim 0.01\,M_\odot$ smaller He cores.

For a further comparison, we constructed two ZA\-EHBs from a He main-sequence star 
with a core mass of $0.47\,M_\odot$ using type I opacities as well as type II.  
Type II opacities differ from type I because of the composition 
change of C and O during thermonuclear evolution.
It is evident from Figure~\ref{dormancomparison} that type I opacities
shift the evolutionary tracks away from the distribution of observed sdB stars.
The significant size of this effect indicates the importance of
the opacities for the position of the ZAEHB.

Our use of OPAL type II opacities, instead of the MESA default type I opacities, appears 
to be the main reason why our EHB evolutionary tracks agree more closely with the observed
points than those of \citet{Ostensen2012}.

The $0.47\,M_\odot$ core mass, type II ZAEHB we constructed 
was also found to be essentially identical with the ZAEHB derived by
Fontaine and collaborators from static models having the same core
mass (Fontaine, priv.\ comm.).  Their models include radiative
levitation as well as gravitational settling, and use opacities
equivalent to OPAL type II in MESA.

In Figure~\ref{bloemencomparison}, we compare our sdB models from Figure~\ref{sdBtracks1}
to models by \citet{Bloemen2014}.
We chose a subset of their models having sdB masses closest to our models.
Theirs all have a total sdB mass of $0.47\,M_\odot$ with slightly different 
He core masses of $M_{\rm{He-core}} = 0.4699, 0.4698, 0.4693, 0.4678\, M_\odot$ 
(from bottom to top).
Since \citet{Hu2011} and \citet{Bloemen2014} focused on the pulsational properties of the sdB stars, they
specifically included radiative levitation in order to recover the iron-group opacity bump. They used
opacities from the Opacity Project \citep{Badnell2005}, which are independent of OPAL type II 
opacities.

The Bloemen et al.\ evolutionary tracks agree remarkably well 
with our tracks computed without radiative levitation, indicating
that this process has little effect on the shapes or the position in the $\log g - T_{\rm{eff}}$ diagram.

We conclude that MESA models with our standard set of input parameters
(Table~\ref{simulationphysics}), are therefore quite capable of reproducing the ZAEHB
and evolutionary tracks of sdB stars 
derived using other stellar codes, as long as the He core masses and opacities are essentially the same. 
The small offset of our sdB tracks relative to the observational points in 
Figure\,\ref{sdBtracks1} could therefore be due to either the evolutionary He core masses or the
opacities.

The canonical timescale for the sdB lifetime is about $100\,\rm{Myr}$ \citep{Dorman1993, Charpinet2000}.
We calculated sdB lifetimes of approximately $140-170\,\rm{Myr}$ for
$M_{\rm{ini}} = 1.0\,M_\odot$ (top part of Table~\ref{1_2Msun_table}), 
in fair agreement with the earlier values, and in very good agreement with \citet{Bloemen2014},
do who found lifetimes of approximately $183, 180, 149, \rm{and} 122~\rm{Myr}$ (from bottom to top)
for the four models shown in Figure\,\ref{bloemencomparison}.
We note that the lifetimes in our simulations are not monotonic functions of mass, but 
depend upon details of the mixing  algorithms, which seem to be very sensitive to the initial conditions(see \S\ref{conv_core_section}).
The smoothest evolutionary tracks are those in which mixing is dominated by atomic diffusion.

Our models show a growing convective core starting from $\sim 0.1\,M_\odot$ at the ZAEHB up to $\sim 0.25\,M_\odot$ 
at core He exhaustion. To compare our results with the convective core masses inferred from asteroseismology, we calculated 
time-averaged convective core masses as listed in Table~\ref{1_2Msun_table}. The $M_{\rm{ini}} = 1.0\,M_\odot$
sequence average core masses are in the range of $0.17-0.18\,M_\odot$ which is substantially smaller than the values found by
\cite{VanGrootel2010a, VanGrootel2010} and \cite{Charpinet2011}.

\subsection{Masses of sdB stars}

According to a study of the empirical mass distribution of sdB stars
\citep{Fontaine2012}, the median value of the sdB mass is $M =
0.471\,M_\odot$, with the range from $0.439 {\rm \ to \ } 0.501\,M_\odot$ containing
$68.3\%$ of the stars.  The maximum masses of our evolutionary sdB models lie
within this range, but they are somewhat smaller than the
median of the observed distribution.

The He core mass on the ZAEHB is determined by the core mass at the
point when the hydrogen shell burning is quenched near the tip of the
RGB, whether it is extinguished prematurely by mass stripping or by
the onset of the He flash. 
Any mass loss occurring between the RGB tip
and the ZAEHB only reduces the amount of residual hydrogen envelope;
uncertainties in the mass loss rates would merely shift the sdB along
the ZAEHB appropriate to its core mass.

We therefore investigated
the effects on the core mass due to the interplay of initial mass,
initial composition and conditions during the He flash.  The results
are compiled in Table~\ref{heflashmasses}.

Larger initial masses than $1.0\,M_\odot$ always produce smaller He cores.
Initial masses smaller than $1.0\,M_\odot$ would produce larger He
cores in the absence of winds, but winds reduce the mass of the 
He core at the He flash for lower mass stars.  If the actual mass loss 
rates on the MS and RGB were much smaller than predicted by the Reimers 
formula as implemented in MESA (with a coefficient of 0.5), initial masses lower
than $0.9\,M_\odot$ could result in slightly larger He core masses.
However, since stars with $M_{\rm{ini}}$ smaller than about
$1.0\,M_\odot$ have not had time to evolve to the RGB tip in the
lifetime of the galactic disk, varying the initial mass of the MESA
progenitor stars does not help to produce larger He cores than we get 
from our standard model.

Next we investigated the effect of different initial compositions.
Table~\ref{heflashmasses} shows that reducing the initial metallicity
to $Z\approx 0.01$ and the helium abundances to $Y\leq 0.26$ would
increase the He core masses to $0.471 - 0.473 M_\odot$,
i.e.\ to the median mass of the empirical mass distribution.
However, since we could not realistically account for He core masses
much higher than $M_{\rm{He-core}}=0.473\,M_\odot$ with composition
changes alone, some additional factor would still be required to
produce model sdB stars corresponding to the upper half of the
observed mass distribution.

\begin{figure}[ht]
 \centering
 \includegraphics[width = 0.47\textwidth]{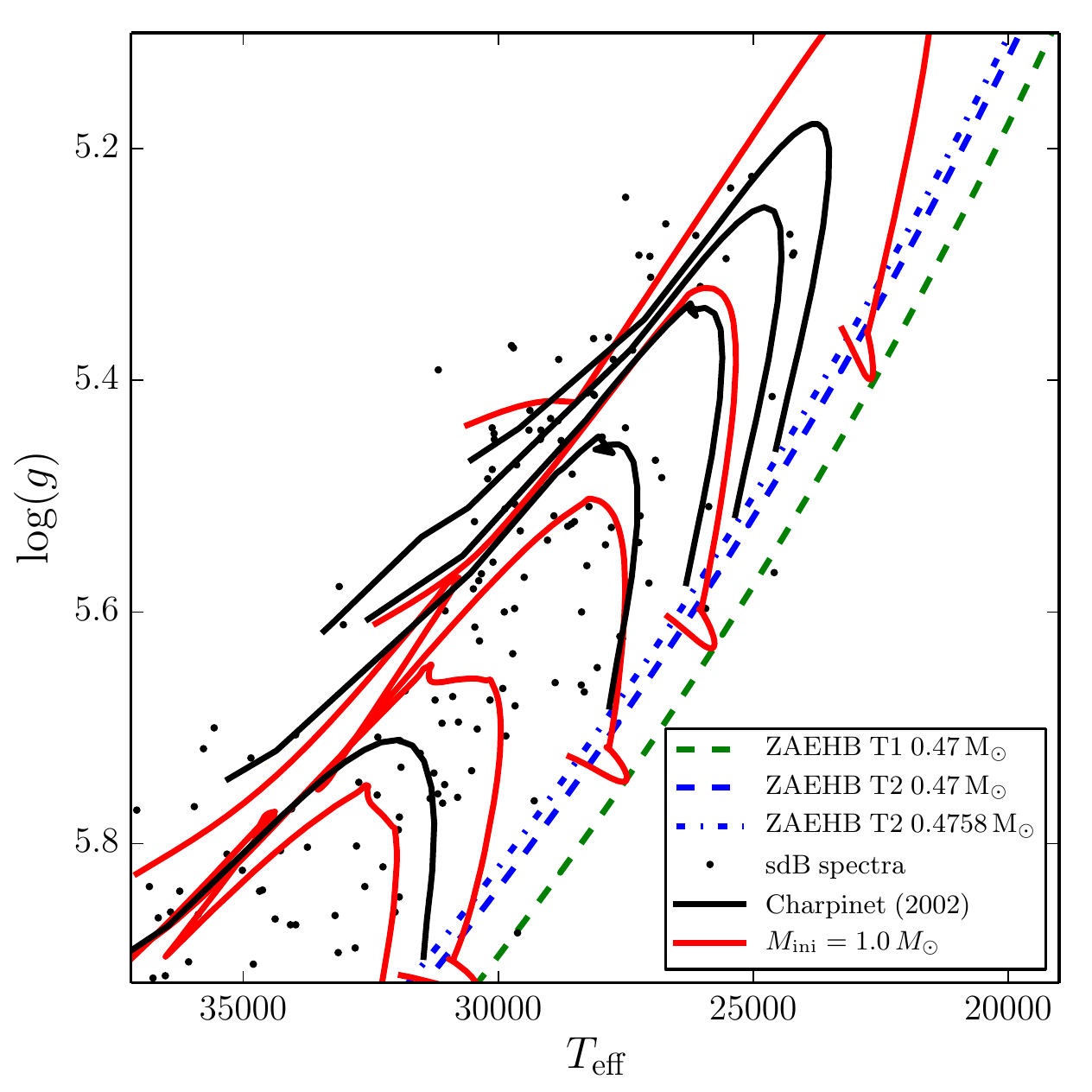} 
 \caption{We compare evolutionary tracks (black solid curves)
   calculated by Ben Dorman with $M_{\rm{He-core}} = 0.4758\,M_\odot$
   \citep{Charpinet2002c} with a MESA ZAEHB constructed using the same
   He-core mass and OPAL type II opacities (dot-dashed line; blue in
   the online journal).  The latter agrees remarkably well with the
   starting points of Dorman's tracks.  For comparison, we show our
   evolutionary tracks from Fig.\,\ref{sdBtracks1} (gray/red solid
   lines; OPAL II opacities).  We also display two MESA ZAEHB's for
   $M_{\rm{He-core}} = 0.47\,M_\odot$, close to the mean of the
   empirical mass distribution of sdB stars \citet{Fontaine2012};
   the gray (green) lower dashed line was constructed with OPAL I 
   opacities and the black (blue) dashed line with OPAL II 
   opacities.  The spectroscopic data points (dots) are the same as in
   Fig.\,\ref{sdBtracks1}.}
 \label{dormancomparison}
\end{figure}

\begin{figure}[ht]
 \centering
 \includegraphics[width = 0.47\textwidth]{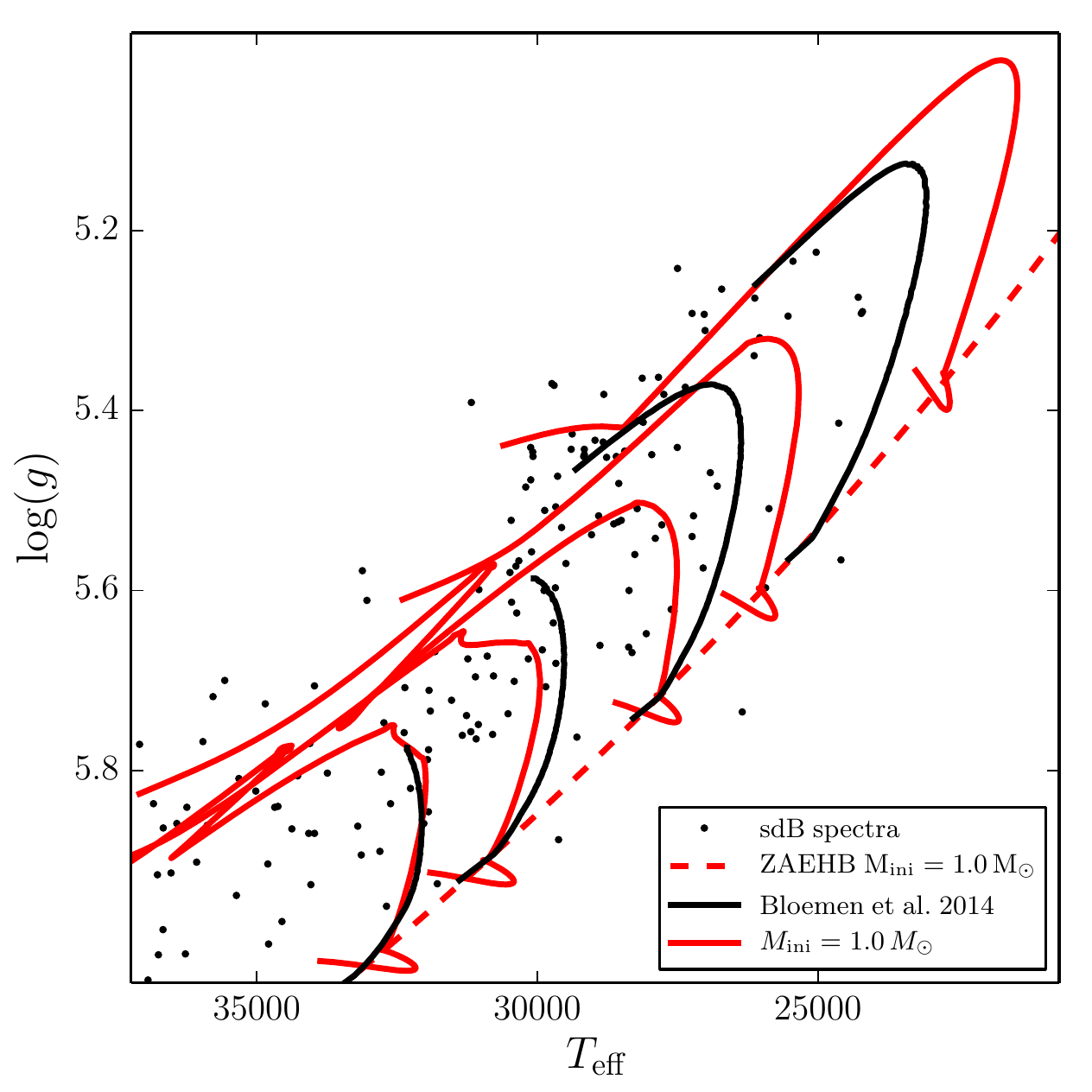} 
 \caption{We compare our evolutionary tracks (gray; red in the online journal) and the 
 corresponding ZAEHB (dashed gray/red ) from Fig.\,\ref{sdBtracks1}
 with selected stellar models from \citet{Bloemen2014} 
 having a total mass of $M_{\rm{sdB}} = 0.47\,M_\odot$ and helium core masses at the ZAEHB 
 of $M_{\rm{He-core}} = 0.4699, 0.4698, 0.4693, 0.4678\, M_\odot$ (from bottom to top).
 The spectroscopic data points (dots) are the same as in Fig.\,\ref{sdBtracks1}.}
 \label{bloemencomparison}
\end{figure}

Even when sdB models are
constructed to have He core masses of $0.470 M_\odot$, or even $0.4758
M_\odot$, the resulting ZAEHB's still appear to be a bit too far to the lower
right in the $\log g - T_{\rm{eff}}$ diagram.  This is true for other 
stellar models as well as ours.   According to the empirical
distribution, a third of the observed points in
Figure\,\ref{dormancomparison} should have evolved from ZAEHB's
corresponding to He core masses between $0.44\, M_\odot$ and $0.47\,
M_\odot$.  Instead, the type II opacity ZAEHB for
$M_{\rm{He-core}}=0.47\,M_\odot$ (upper/blue dashed line) 
appears to fall significantly below the lower envelope of the vast majority
of observed points. Similarly, the upper envelope of observed
points in Figures~\ref{sdBtracks1} and \ref{bloemencomparison} seems overpopulated,
especially since the evolution is much faster as the stars begin to exhaust the He in their
cores.

Mixing processes in the progenitor star do influence the He core mass up to
the onset of the He flash but the effect is small with current stellar evolutionary codes.  
We ran a few test models without atomic diffusion,
starting from the pre-main-sequence, and compared them
to our standard progenitor model with atomic diffusion. As seen previously by 
\citet{mrr07} and \citet{Hu2008}, adopting atomic diffusion produces only 
a marginal increase in the He core mass of $\sim 0.0015\,M_\odot$ at the RGB tip,
and consequently in the ZAEHB He core mass.

In principle the He core mass might be increased prior to the He flash
by mixing processes, but within the present framework this is an
arbitrary and not particularly helpful addition. For example, if convective overshoot is active
during the progenitor evolution, the He core mass is enhanced by $\sim
0.0005\,M_\odot$ for a overshoot parameter of $f_{\rm{ov}} = 0.02$, which 
is a change too small to be of significant benefit.

\begin{table*}[ht]
 \centering
 \caption{He-core masses for a variety of sdB models}
\begin{tabular}{ccccccc}
 \tableline
\tableline
\\
$M_{\rm{ini}}$ & $Z_{\rm{ini}}$ & $Y_{\rm{ini}}$ & $X_{\rm{ini}}$ & $M_{\rm{He-core}}$ & Age$_{\rm{He-flash}}$ & Comment\\
$[M_\odot]$ &  &  &  & $[M_\odot]$ & [$10^9\,\rm{yrs}$] & \\
\\
\tableline
\\
2.0 & 0.02 & 0.28 & 0.70 & 0.436 & 1.085 & \\
1.2 & 0.02 & 0.28 & 0.70 & 0.464 & 6.212 & \\
1.0 & 0.02 & 0.28 & 0.70 & 0.466 & 11.896 & ``standard model'' \\
0.9 & 0.02 & 0.28 & 0.70 & 0.467 & 17.142 & \\
0.8 & 0.02 & 0.28 & 0.70 & 0.461 & 25.588 & \\
0.7 & 0.02 & 0.28 & 0.70 & \multicolumn{2}{c}{no He-flash}  & \\
\\
\tableline
\\
1.0 & 0.02 & 0.28 & 0.70 & 0.466 & 11.848 & no winds\\
0.9 & 0.02 & 0.28 & 0.70 & 0.467 & 17.075 & no winds\\
0.7 & 0.02 & 0.28 & 0.70 & 0.472 & 39.514 & no winds\\
\\
\tableline
\\
1.0 & 0.025 & 0.30 & 0.675 & 0.461 & 11.502 \\
1.0 & 0.02 & 0.25 & 0.73 & 0.471 & 14.529 & \\
1.0 & 0.02 & 0.35 & 0.63 & 0.453 &  7.293 & \\
1.0 & 0.01 & 0.26 & 0.73 & 0.471 & 9.976 & \\
1.0 & 0.01 & 0.25 & 0.74 & 0.473 & 10.461 & \\
1.0 & 0.005 & 0.25 &0.745 & 0.473 & 7.999 & \\
\\
\tableline
\\
\multicolumn{7}{c}{
\begin{minipage}{1.2\columnwidth}
\footnotesize \textbf{Notes.} The He-flash mass is approximated by the mass coordinate where the hydrogen mass fraction per cell falls below X=0.1 right before the He-flash. \end{minipage} }
\end{tabular}\\

\label{heflashmasses}
\end{table*}

\subsection{Observational constraints on initial masses and metallicities} \label{obs constraints}

Our choice of a $1.0\,M_\odot$, solar composition progenitor for most
of our ZAEHB models was based on available data for sdB stars in old
open clusters.  Unfortunately, it is not possible to derive ages and
progenitor masses for individual field sdB stars, and their current
atmospheric compositions are completely independent of their original
metallicities due to strong diffusion in their extremely thin
envelopes. In general, since most well-studied (e.g.\ nearby) sdB
stars belong to the Galaxy's old disk population
\citep{1991PhDT........10S}, all we know is that their progenitors
must have been low mass stars less than about 10 Gyr old with
metallicities greater than about 1/10 solar.  More precise estimates
of initial masses and metallicities can be obtained for sdB stars
that are members of clusters, but only two open clusters are
known to contain hot subdwarfs: NGC\,6791 (5--6 sdB stars) and
NGC\,188 (one sdB).  These also happen to be two of the oldest open
clusters known, 8.3 Gyr and 6.2 Gyr, respectively \citep{Brogaard2012,
  Meibom2009}, as well as two of the most metal rich, with 2.5 times solar
metallicity (Z = 0.05) and slightly greater than solar metallicity (Z
= 0.026) \citep{Heiter2014}.  NGC\,6791 is a much more populous
cluster, having four times as many normal He core burning red
giant clump stars as NGC\,188. The cluster turnoff masses, derived
very precisely from eclipsing binaries, are 1.087 and 1.103 $M_\odot$,
respectively; the ZAMS masses of the currently observed sdB
progenitors would have been a few hundredths of a solar mass larger.

It turns out that there are no other well-studied clusters with 
ages $\gtrsim$ 6 Gyr and supersolar metallicities.  However, 
it is possible to compare the statistics of sdB stars in clusters that 
are comparably old and more metal-poor, or somewhat younger and comparably 
metal-rich, or both younger and more metal-poor.  While few open clusters
have been definitively searched for hot stars at ultraviolet
wavelengths \citep[e.g.\,][]{Carraro2013, Zloczewski2007}, the
presence or absence of sdB's is obvious in many open cluster
color-magnitude diagrams (CMD's), wherever the blue edge of the field
star distribution is significantly redder than the colors of hot
subdwarfs.  (Note that here we are concerned only with He
core burning EHB stars, not fainter cataclysmic variables or other
somewhat cooler objects sometimes suggested to be EHB candidates.)

We conducted a literature search of all well-studied old disk clusters
having sufficiently deep CCD photometry to reveal faint sdB candidates
and populous enough to have a distinct red giant clump (as a measure
of the relative size of the cluster sample) and found the following
results.
There are no EHB stars in the CMD's of four old open clusters (Be\,17,
Be\,32, Be\,39, and Cr\,261) with comparable ages to NGC\,6791 and
NGC\,188 (5.5 to 9 Gyrs) and lower metallicities ($0.004 < $Z $<
0.015$).  The combined red giant clump population in the CMD's of
these four clusters is slightly larger than the red clump in NGC\,6791
\citep{Bragaglia2006, Tosi2007, Bragaglia2012, Gozzoli1996}.  The CMD's
of three other clusters (NGC\,6819, NGC\,6253, and NGC\,7142) with
similar metallicities to NGC\,6791 and NGC\,188 ($0.025 <$ Z $< 0.05$)
and younger ages (3 to 4 Gyr) have about the same total number of
red giant clump stars as NGC\,6791 and also have no EHB stars
\citep{Jeffries2013, Kaluzny2014, Sandquist2013}.  A sample of eight
younger and more metal-poor clusters (Be\,22, Be\,31, Be\,66,
Mel\,66; Tr\,5, Be\,29, M\,67, NGC\,224; 2.5 to 5 Gyr, $0.003 <$ Z $<
0.006$)\citep{Fabrizio2005, Cignoni2011, Andreuzzi2011, Carraro2014,
  Kaluzny1998, Tosi2004, Montgomery1993, Kaluzny2006} with a
combined red clump population more than twice that of NGC\,6791,
contains a total of one, still unconfirmed, EHB candidate (in Mel\,66,
3.4 Gyr, Z = 0.01; \citet{Zloczewski2007}).

To summarize, all of the confirmed sdB members of old disk clusters
are found in NGC\,6791 and NGC\,188, both of which have supersolar
metallicities and are older than 6 Gyr.  15 old open clusters either
somewhat younger and/or more metal-poor than NGC\,6791 and NGC\,188
have produced, at most, a single sdB star, instead of the 20 or more
such stars that would be expected if the fraction of hot subdwarfs was
similar in all old disk clusters. The open cluster data suggest
that lower mass, very metal-rich progenitors produce a much higher
fraction of field sdB stars.


Table~\ref{heflashmasses} shows that the initial masses inferred for
the known sdB members of old disk clusters, $1.1$ to $1.2\,M_\odot$
are consistent with observations of field sdB stars in the sense that
they correspond to nearly the largest possible values of the helium
core mass that could have been produced in the lifetime of the
galactic disk (even if the latter are not large enough to match the
observations).  Significantly lower initial helium, and perhaps lower
metallicities, can produce slightly larger core masses by the time
of the He flash, but such low abundances are incompatible with the
statistics of sdB cluster members presented above.

Most importantly, our models were evolved almost to the onset of the He
flash before the envelope stripping was initiated.  Assuming that sdB
stars lose the last of their envelopes just as the He core flash is
about to begin requires an unrealistic fine-tuning of the initial
binary parameters.  Instead, most sdB progenitors are expected
to leave the RGB somewhat before the He core flash, when their He core
masses are significantly smaller. \citet{Castellani1993} and \citet{D'Cruz1996} calculated that
stars near the red giant tip with He core masses up to $\sim
0.02\,M_\odot$ less than the He flash core mass can lose their
envelopes and still become EHB stars.

Although the He core masses of our standard MESA sdB models are
already too small compared to observations, it is clear that more
realistic assumptions would reduce the ZAEHB He core masses even
further, increasing the discrepancy with observations.

\subsection{A solution from nuclear astrophysics?}

It might be thought that our inability to produce higher He core
masses is related to the net rates of the helium burning reactions
which are used as defaults in MESA, indicating that these rates might
need revision.

The mass at degenerate ignition is essentially a measure of the peak
temperature, and insensitive to other parameters. In turn, the
temperature at thermal runaway is sensitive to the effective reaction
rate, including both nuclear and electron screening effects.
\citet{Hoyle1954} used this physics plus the core mass-luminosity
relation to infer the existence of an excited state in $^{12}\rm{C}$;  see \S8.1,
\citet{1996sunu.book.....A}, for details. 
We expect a slightly lower effective rate to give a later helium flash,
allowing the core to grow larger.
\citet{Iliadis2007}(\S5.3.1) has reviewed the experimental
situation regarding the rates of helium burning reactions. The
triple-alpha reaction is not directly measured, but is thought to be
reliably estimated ($\pm 35\%$) by indirect means for normal, non-degenerate
conditions. The $\rm^{12}C(\alpha,\gamma) ^{16}O$ reaction is
notorious for its experimental difficulty, but must track the triple-alpha rate 
to avoid significant nucleosynthesis consequences.

To test the sensitivity of core mass to reaction rate, we performed 
a numerical experiment in which we simply
lowered the effective nuclear reaction rates for our stellar
progenitor model.  We decreased the effective nuclear reaction rates
for the triple-$\alpha$ process and the nitrogen reactions by a factor
of four, which led to a He-core mass of $0.473\,M_\odot$ (factor of
two: $0.469\,M_\odot$) for the $M_{\rm{ini}}=1.0\,M_\odot$ model,
bringing the He-core mass slightly above the median of the sdB mass
distribution. 
These arbitrary changes are three times the estimated $3\sigma$ error, 
and seem to be an very unlikely resolution of the discrepancy in He core 
masses of sdB stars.

\section{CONVECTIVE CORES OF SDB STARS} \label{conv_core_section}

Using our standard set of input physics, including the well described 
set of physical processes causing atomic diffusion, we found that our 
time-averaged convective core masses are too small in comparison to 
the data inferred from asteroseismology. 
A combination of type II opacities plus atomic diffusion is clearly
necessary, although insufficient.  A further increase of the
average convective core mass of about $ 0.04-0.09\,M_\odot$ is
needed for agreement with the values found by \citet{VanGrootel2010a,
VanGrootel2010} and \citet{Charpinet2011}.

The convective core size is mainly determined by the convective and
diffusive mixing processes during the evolution of the sdB star. For our
standard models, we used the Schwarzschild criterion to determine
the convective boundary and the MLT picture of convection.

In the following sections, we further investigate the different
input physics options available in MESA to illustrate how the
interplay of opacities, diffusion and overshoot effect
the convective core growth, and thus the sdB lifetime, as well as the
position in the $\log g - T_{\rm{eff}}$ diagram.

\subsection{Direct comparison of different input physics}

\begin{figure*}[ht]
 \centering
 \includegraphics[width = \textwidth]{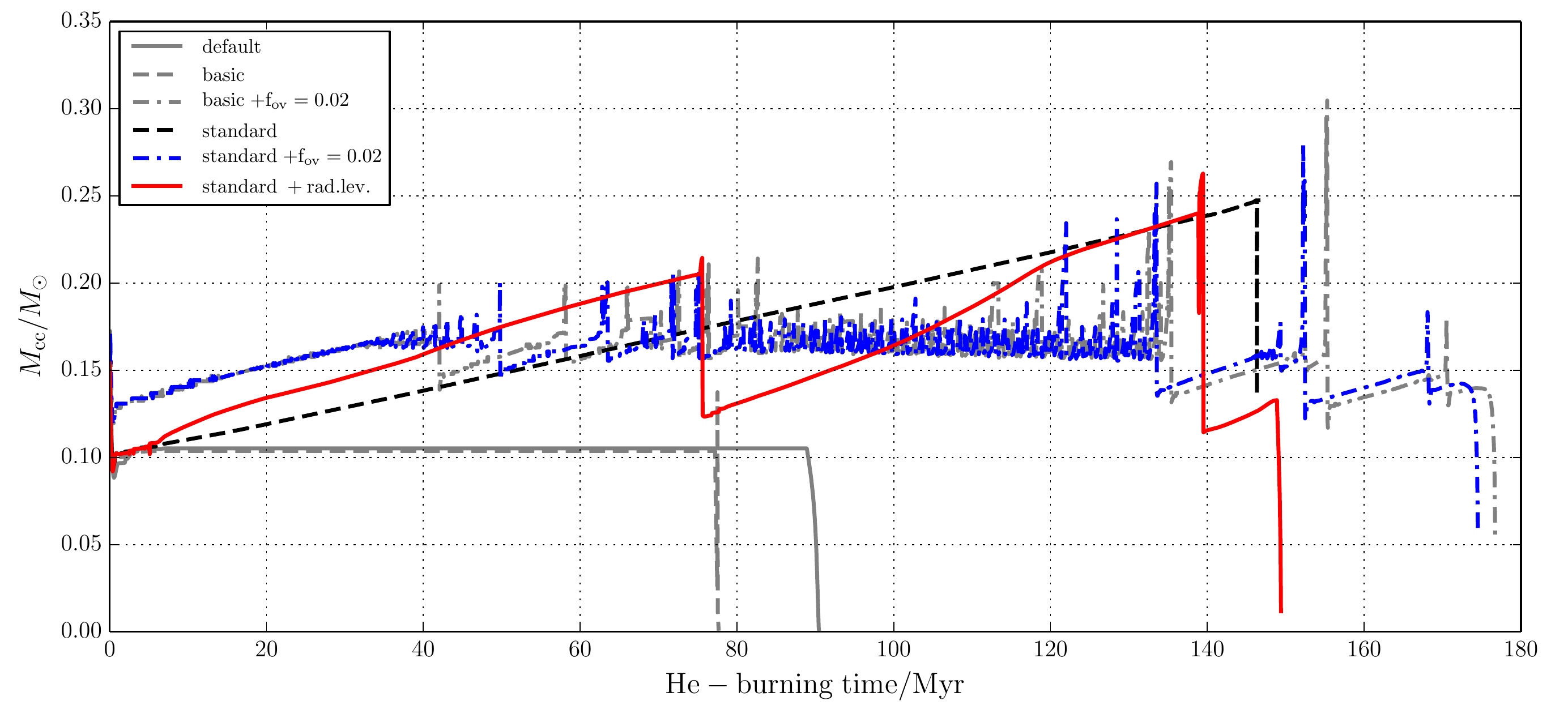}
 \caption{Convective cores as a function of time, with different
   input physics.  The largest cores result from a 
   combination of type II opacities and either overshoot or atomic diffusion, 
   which conspire to induce core growth. Large  values for the 
   overshoot parameter do not produce core growth, just larger
   cores.  The rapid fluctuations in the models with overshoot suggest that 
   the mixing algorithm experiences numerical instabilities at the boundary. 
   The model with radiative levitation shows a sawtooth curve that we also find 
   in some of our other standard models without radiative levitation. 
   The sawtooth is caused by the convection zone growing and collapsing periodically.} 
 \label{physics_conv_core}
\end{figure*}

\begin{figure}[ht]
 \centering
 \includegraphics[width = 0.47\textwidth]{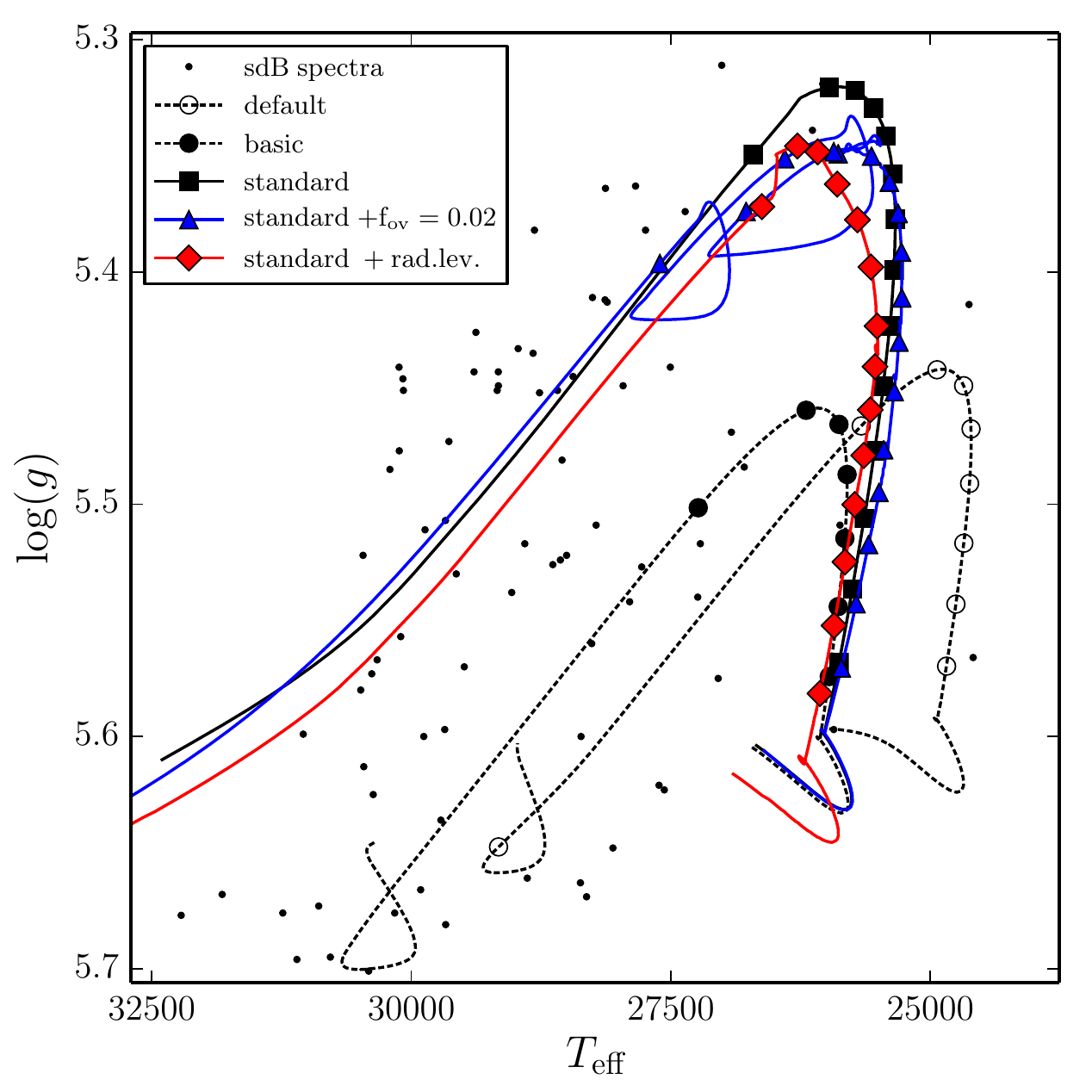}
 \caption{Evolutionary tracks starting from the same standard initial
   model, but with different input physics.  The sdB observational
   data are shown as single symbols and the evolutionary tracks are
   shown as symbols connected by lines. (The basic model with
   overshoot was not plotted, since it is indistinguishable from the
   standard model with overshoot.)  As in
   Fig.\,\ref{physics_conv_core}, the most significant change results
   from core growth due to the combination of type II opacities and
    either overshoot or atomic diffusion.} 
 \label{physics_logg_Teff}
\end{figure}

 We compared six different sdB models, all having the same
standard $M_{\rm{ini}}=1.0\,M_\odot$ progenitor model and
$M_{\rm{new}}=0.48\,M_\odot$,  using different options for the
input physics as specified in Table~\ref{physics_conv_cores_table}.  
The default model uses most of the MESA
default input physics, including OPAL type I opacities and no atomic
diffusion.  The basic model uses the OPAL type II opacities and again
no atomic diffusion. The third model in this comparison adds
convective overshoot to the basic model.  Our standard
sdB model has the input physics summarized in
Table~\ref{simulationphysics}.   The fifth model adds convective 
overshoot to the standard model, and the last model shows the effect
of radiative levitation.\footnote {Radiative levitation in this model
makes use of OPAL Project opacities \citep{Badnell2005}, which include 
additional data for iron group elements.}

The results for the convective core masses are shown in
Figure\,\ref{physics_conv_core}.  The default and the basic model both
have a constant convective core mass of $0.1\,M_\odot$ throughout the
sdB lifetime, which is therefore significantly shorter than for the standard
model.  Adding convective overshoot with $f_{\rm{ov}} = 0.02$ to the
basic model allows the convective core to grow from a value of
about $0.13\,M_\odot$ to $0.16\,M_\odot$.  From then on, the convective core 
size in the basic +$f_{\rm{ov}}$ model remains roughly constant, 
but the convective boundary shows rapid irregular fluctuations probably 
related to inadequacies in
the formulation of the overshoot algorithm. The sdB lifetime for this model is
extended by about a factor of two compared to the default and the
basic model.

The standard model also produces a steadily growing convective core up
to a mass of $\approx 0.25\,M_\odot$ at He core exhaustion, as
described in \S\,\ref{model_comparison_section}.  This is our only 
model in which the increase in the convective core mass is monotonic 
and smooth. The sdB lifetime is $\sim 50\,\rm{Myr}$ longer than for the
default and basic models, in full agreement with the results of
\cite{Bloemen2014}. 

\begin{table}[h]
\centering
\caption{Global properties for sdB models \\with different input physics}
\label{physics_conv_cores_table}
\begin{tabular}{ccc}
\tableline
\tableline
\\
Model	& SdB lifetime	& Average \\
physics & $[10^6\,\rm{yrs}]$ & $M_{cc} [M_\odot]$ \\
 \\
\tableline
\\
default~ (OPAL I, no diffusion) &	90.7	&	0.105	\\
basic~ (OPAL II, no diffusion)  &	77.8	&	0.105	\\
basic + $f_{\rm{ov}} = 0.02$	&	176.8	&	0.157	\\
standard~ (OPAL II, with diffusion) &	146.8	&	0.172	\\
standard + $f_{\rm{ov}} = 0.02$	&	174.6	&	0.157	\\
standard + rad. levitation      &	146.9	&	0.168	\\

\\
\tableline
\end{tabular}
\end{table}

A default model with convective overshoot (not listed in the table) showed
that the current MESA overshoot algorithm simply forces the size of the
convective core to have a larger constant value, whereas the combined 
effect of using type II
opacities and either overshoot (basic + $f_{\rm{ov}}$) or atomic diffusion 
(standard) is to cause core growth.

When overshoot is added to the standard model, we see the same
behavior discussed above for the basic model with overshoot, including
the same rapid fluctuations, except that the maximum convective core size 
and the lifetime are both larger.

Adding radiative levitation slightly decreases the average convective
core mass of our standard model because of the sawtooth behaviour seen
in Figure\,\ref{physics_conv_core}. It has a negligible effect on the
sdB lifetime.  We encountered the sawtooth behavior in many of our other
standard models (i.e.\ without radiative levitation). The convection
zone grows and collapses periodically in these models.

We believe this not a numerical artefact but rather a consequence of
the model physics.  The effect is reminiscent of He-flash models
  by \citet{Mocak2008,Mocak2009}.  It seems to be strongly
dependent on the different model parameters, especially the new mass
after stripping off the envelope.  This behavior changes the profile
of the helium, carbon and oxygen abundances as a function of time,
compared to a smooth curve, although it does not help to explain the
discrepancies between the convective core masses of our models and the
ones inferred from asteroseismology.

Figure\,\ref{physics_logg_Teff} shows the evolutionary tracks of five
of the six models compared to sdB observational data.  When type II
opacities are used (basic model; black circles) instead of type I
opacities (default model; open circles) the evolutionary paths shift
to higher temperatures and very slightly higher gravities.  The most
significant difference is the evolution to much lower gravities caused
by the convective core growth, due to the combination of type II
opacities and either overshoot in the basic model or diffusion in the
standard model (black squares).  The resulting tracks extend high
enough to explain a majority of the observed points. The standard
model plus radiative levitation (gray diamonds; red in the online
version) gives a similar result.  Adding overshoot to either the basic
model (not plotted) or the standard model (gray squares; blue in the
online version) also turns out fairly similar, except for the suspicious
looping behavior toward the end of the He core burning
in Figure\,\ref{physics_logg_Teff}, corresponding to the large spikes in
Figure\,\ref{physics_conv_core}.  (The tracks of the two models with
convective overshoot are so similar that for clarity we only show the
standard model with overshoot.)

\begin{table}[ht]
\centering
\caption{Properties for sdB models with additional overshoot.}
\label{convectivecores}
\begin{tabular}{cccccc}
\tableline
\tableline
\\
 &  \multicolumn{2}{ c }{Standard model}& \multicolumn{2}{ c }{Basic model}\\ 
 &  \multicolumn{2}{ c }{(OPAL II and diffusion)}& \multicolumn{2}{ c }{(OPAL II, no diffusion)}\\ 
 \\
\tableline
\\
overshoot	& sdB 	& Convective & sdB 	& Convective \\
 parameter& lifetime  & core mass &  lifetime & core mass\\
 $f_{\rm{ov}}$ & $[10^6\,\rm{yrs}]$ & $M_{cc} [M_\odot]$ & $[10^6\,\rm{yrs}]$ & $M_{cc} [M_\odot]$ \\
 \\
\tableline
\\
no overshoot &	146.8 &	0.172 & 77.8 & 0.105\\
0.01 &	 194.8 &	0.149 & 179.4 & 0.152\\
0.02 &	 174.6 &	0.157 & 176.8 & 0.157\\
0.04 &	 176.2 &	0.179 & 174.98 & 0.179\\
0.08 &	 170.3 &	0.227 & 170.2 & 0.227 \\
0.10 &	 179.3 &	0.252 & 179.2 & 0.252\\
\\
\tableline
\end{tabular}
\end{table}

\subsection{Achieving larger convective cores with convective overshoot\label{overshoot}}

Although we see erratic behavior when convective
overshoot is added to any model computed with type II opacities, we nonetheless
calculated a few models with varying overshoot parameters to
investigate whether a sufficiently large overshoot parameter in the current
implementation would bridge the discrepancy in convective core size
between our stellar models and asteroseismology.

Table~\ref{convectivecores} lists the results of two model sequences with 
varying amounts of convective overshoot, as implemented in MESA.
 Except for the value of the overshoot parameter, these model sequences 
are the same as the basic and standard 
models in Table~\ref{physics_conv_cores_table}.  They start from the same progenitor model evolved 
with standard input physics ($M_{\rm{ini}} = 1.0\,M_\odot$, $M_{\rm{new}} = 0.48\,M_\odot$).  

An increase in the overshoot parameter, $f_{\rm{ov}}$, directly affects the
extent of the convective core.  To achieve convective cores as large as those determined by
\citet{VanGrootel2010a, VanGrootel2010} and \citet{Charpinet2011}, 
the overshoot parameter must be greater than $0.08$. 
For reference, we note that overshoot parameters
$f_{\rm{ov}}\sim10^{-5}{\rm\ to\ }10^{-4}$ have been used for $3\,M_\odot$ stars
and overshoot parameters $f_{\rm{ov}}~\sim 10^{-3}{\rm\ to\ }10^{-2}$ for
$1.5\,M_\odot$ stars \citep{Paxton2013}. \citet{Herwig2000} suggested
$f_{\rm{ov}}\sim 0.016$ for stellar interiors in order to reproduce
the models of \citet{Schaller1992}.

We have to increase the overshoot parameter by
at least a factor of five to reproduce values close to the
results of \citeauthor{VanGrootel2010} and
\citeauthor{Charpinet2011} even in the presence of atomic diffusion in our standard model.
The fact that the overshoot parameter must be varied by orders of magnitude 
in different types of stars underscores the physical inadequacy of 
the current treatment of overshooting.

It is interesting that convective overshoot seems to hinder the
convective core growth that occurs as a result of atomic diffusion and
type II opacities in our standard model. Table~\ref{convectivecores} shows
that overshoot parameters of $0.01-0.02$ produce smaller average convective core masses 
than the standard model with no overshoot.  Larger convective core masses are achieved only for
$f_{\rm{ov}} \geq 0.04$.  It appears that overshoot
interacts in an unexpected way with diffusion as well as with type II opacities.

The overshoot algorithm increases core sizes by increasing the importance of the numerical diffusion term \citep{Eggleton1972}, which in turn flattens the composition gradients. Smaller gradients make atomic diffusion weaker; actually, overshoot should enhance atomic diffusion. 

\section{CONCLUSION} \label{conclusion}

MESA is an excellent software environment to explore implications of
numerical experiments, and to find inadequacies in our theoretical
ideas about how stars evolve.

%
%
Our MESA sdB models reproduce the general properties of the ZAEHB and
the characteristic hook shape of the helium core burning evolutionary
path.  We have demonstrated that MESA is fully capable of 
recreating previous theoretical EHB results, e.g. the position of the tracks in the 
$\log g - T_{\rm{eff}}$ plane and the model time scales
\citep{Charpinet2002c, Bloemen2014}, 
as long as we use Type II opacities and
start with the same value for the He-core mass on the ZAEHB.  

However, although we are able to produce structures which are consistent with the asteroseismology of sdB stars, {\em we cannot evolve to these conditions with plausible parameters for standard stellar evolution.} Our largest total sdB masses are smaller than the median mass of the empirical sdB mass distribution.
More importantly, the computed helium burning cores are smaller than inferred by observation. This is an error in convective mixing in the deep interior, far from any superadiabatic region in the envelope. It cannot be blamed on MLT alone, and is {\em likely to be related to the treatment of the convective boundary.} 

A clue for the solution of this problem comes from 3D simulations of stellar convection having sufficient resolution to show turbulent flow \citep{Viallet2013}. These simulations provide a closure for the Reynolds averaged Navier-Stokes (RANS) analysis, which in turn suggests approximations which would allow this behavior to be implemented in a stellar evolutionary code. They do not require calibration to astronomical data. See \cite{amvcl15} for details; here we summarize a few features relevant to the sdB problem. A robust feature of these simulations is the development of a boundary layer between convective and nonconvective regions, in which the radial flow is turned so that convection stays within the convective region. This braking layer is subadiabatic and well mixed, so that mixing extends beyond the radius defined by the Schwarzschild criterion; ''extra mixing" is required, and not an option. The braking layers are narrow, with a width which depends upon the stiffness of the stable layer and the velocity of the flow; stiff  boundaries and slow flow have narrower braking layers. The bulk of the convective region (away from the convective boundary) is moderately well described by MLT, a fact which, along with the free parameter, explains its level of success and enduring popularity. 

It appears that convection as presently implemented in stellar evolution codes is inadequately accurate for precise tests, such as those imposed by asteroseismology. We also note that such a boundary modification would tend to shift the standard solar model toward the Asplund abundances \citep{Asplund2009}. The inferred core sizes of the sdB stars and the 3D simulations suggest a consistent picture may be obtained with introduction of more physically consistent boundary conditions into stellar evolutionary codes.

\acknowledgments
This work was supported in part by NSF Award 1107445 at the University
of Arizona. We thank St\'{e}phane Charpinet, Gilles Fontaine, Conny
Aerts, Steven Bloemen and John Lattanzio for helpful and constructive discussions,
and  G. Bono for interesting questions.


\begin{thebibliography}{97}
\expandafter\ifx\csname natexlab\endcsname\relax\def\natexlab#1{#1}\fi

\bibitem[{Andreuzzi {et~al.}(2011)Andreuzzi, Bragaglia, Tosi, \&
  Marconi}]{Andreuzzi2011}
Andreuzzi, G., Bragaglia, A., Tosi, M., \& Marconi, G. 2011, \mnras, 412, 1265

\bibitem[{Arnett(1996)}]{1996sunu.book.....A}
Arnett, D. 1996, {Supernovae and Nucleosynthesis} (Princeton University Press,
  1996)

\bibitem[{Arnett \& Meakin(2011)}]{Arnett2011}
Arnett, D.~W., \& Meakin, C. 2011, \apj, 741, 33

\bibitem[{Arnett {et~al.}(2015)}]{amvcl15} Arnett, W. D., Meakin, C., Viallet, M., Campbell, S., and Lattanzio, J.,
2015, in prep.

\bibitem[{Asplund {et~al.}(2009)Asplund, Grevesse, Sauval, \& Scott}]{Asplund2009}
Asplund, M., Grevesse, N., Sauval, A.~J., \& Scott, P. 2009, \araa, 47, 481

\bibitem[{Badnell {et~al.}(2005)Badnell, Bautista, Butler, Delahaye, Mendoza, 
Palmeri, Zeippen, Seaton}]{Badnell2005} 
Badnell, N.~R., Bautista, M.~A., Butler, K., {et~al.} 2005, \mnras, 360, 458 

\bibitem[{Baran {et~al.}(2012)Baran, Reed, Stello, \O~stensen, Telting,
  Pak\v{s}tienė, O'Toole, Silvotti, Degroote, Bloemen, Hu, {Van Grootel},
  Clarke, {Van Cleve}, Thompson, \& Kawaler}]{Baran2012}
Baran, A.~S., Reed, M.~D., Stello, D., {et~al.} 2012, \mnras, 424, 2686

\bibitem[{Bergeron {et~al.}(1988)Bergeron, Wesemael, Michaud, \& Fontaine}]{Bergeron1988} 
Bergeron, P., Wesemael, F., Michaud, G., \& Fontaine, G.\ 1988, \apj, 332, 964 

\bibitem[{Bl\"{o}cker(1995)}]{Blocker1995}
Bl\"{o}cker, T. 1995, \aap, 297, 727

\bibitem[{Bloemen {et~al.}(2014)Bloemen, Hu, Aerts, Dupret, \O~stensen, \&
  Degroote}]{Bloemen2014}
Bloemen, S., Hu, H., Aerts, C., {et~al.} 2014, \aap, 569, A123

\bibitem[{B\"{o}hm-Vitense(1958)}]{Bohm-Vitense1958}
B\"{o}hm-Vitense, E. 1958, Zeitschrift fur Astrophysik, 46, 108

\bibitem[{Bragaglia {et~al.}(2012)Bragaglia, Gratton, Carretta, D’Orazi,
  Sneden, \& Lucatello}]{Bragaglia2012}
Bragaglia, A., Gratton, R.~G., Carretta, E., {et~al.} 2012, \aap, 548, A122

\bibitem[{Bragaglia {et~al.}(2006)Bragaglia, Tosi, Carretta, Gratton, Marconi,
  \& Pompei}]{Bragaglia2006}
Bragaglia, A., Tosi, M., Carretta, E., {et~al.} 2006, \mnras, 1502,
  060120032526005

\bibitem[{Bressan {et~al.}(2014)Bressan, Girardi, Marigo, Rosenfield, \&
  Tang}]{Bressan2014}
Bressan, A., Girardi, L., Marigo, P., Rosenfield, P., \& Tang, J. 2014,
  arxiv:1409.2268

\bibitem[{Brogaard {et~al.}(2012)Brogaard, VandenBerg, Bruntt, Grundahl,
  Frandsen, Bedin, Milone, Dotter, Feiden, Stetson, Sandquist, Miglio, Stello,
  \& Jessen-Hansen}]{Brogaard2012}
Brogaard, K., VandenBerg, D.~A., Bruntt, H., {et~al.} 2012, \aap, 543, A106

\bibitem[{Carraro {et~al.}(2013)Carraro, Buzzoni, Bertone, \&
  Buson}]{Carraro2013}
Carraro, G., Buzzoni, A., Bertone, E., \& Buson, L. 2013, \aj, 146, 128

\bibitem[{Carraro {et~al.}(2014)Carraro, de~Silva, Monaco, Milone, \&
  Mateluna}]{Carraro2014}
Carraro, G., de~Silva, G., Monaco, L., Milone, A.~P., \& Mateluna, R. 2014,
  \aap, 566, A39

\bibitem[Cassisi {et~al.}(2007) Cassisi, Potekhin, Pietrinferni, Catelan \& Salaris]{Cassisi2007} {Cassisi}, S., {Potekhin}, A.~Y., {Pietrinferni}, A., 
	{Catelan}, M., {Salaris}, M., 2007, \apj, 661,1094  
  

\bibitem[{Castellani \& Castellani(1993)}]{Castellani1993}
Castellani, M., \& Castellani, V. 1993, \apj, 407, 649


\bibitem[{Chandrasekhar(1961)}]{Chandrasekhar1961}
Chandrasekhar, S. 1961, {Hydrodynamic and Hydromagnetic Stability} (Oxford
  University Press, 1961)

\bibitem[{{Charpinet} {et~al.}(1997){Charpinet}, {Fontaine}, {Brassard},
  {Chayer}, {Rogers}, {Iglesias}, \& {Dorman}}]{Charpinet1997}
{Charpinet}, S., {Fontaine}, G., {Brassard}, P., {et~al.} 1997, \apjl, 483,
  L123

\bibitem[{Charpinet {et~al.}(2000)Charpinet, Fontaine, Brassard, \&
  Dorman}]{Charpinet2000}
Charpinet, S., Fontaine, G., Brassard, P., \& Dorman, B. 2000, \apjs, 131, 223

\bibitem[{Charpinet {et~al.}(2002)Charpinet, Fontaine, Brassard, \&
  Dorman}]{Charpinet2002c}
---. 2002, \apj, 140, 469

\bibitem[{Charpinet {et~al.}(2008)Charpinet, {Van Grootel}, Reese, Fontaine,
  Green, Brassard, \& Chayer}]{Charpinet2008}
Charpinet, S., {Van Grootel}, V., Reese, D., {et~al.} 2008, \aap, 489, 377

\bibitem[{Charpinet {et~al.}(2011{\natexlab{a}})Charpinet, Fontaine, Brassard,
  Green, {Van Grootel}, Randall, Silvotti, Baran, \O~stensen, Kawaler, \&
  Telting}]{Charpinet2011a}
Charpinet, S., Fontaine, G., Brassard, P., {et~al.} 2011{\natexlab{a}}, \nat,
  480, 496

\bibitem[{Charpinet {et~al.}(2011{\natexlab{b}})Charpinet, {Van Grootel},
  Fontaine, Green, Brassard, Randall, Silvotti, \O~stensen, Kjeldsen,
  Christensen-Dalsgaard, Kawaler, Clarke, Li, \& Wohler}]{Charpinet2011}
Charpinet, S., {Van Grootel}, V., Fontaine, G., {et~al.} 2011{\natexlab{b}},
  \aap, 530, A3

\bibitem[{Chayer {et~al.}(2004)}]{Chayer2004} 
Chayer, P., Fontaine, G., Fontaine, M., {et~al.} 2004, \apss, 291, 359 

\bibitem[{Cignoni {et~al.}(2011)Cignoni, Beccari, Bragaglia, \&
  Tosi}]{Cignoni2011}
Cignoni, M., Beccari, G., Bragaglia, A., \& Tosi, M. 2011, \mnras, 416, 1077

\bibitem[{Cox \& Giuli(1968)}]{cox1968principles}
Cox, J.~P., \& Giuli, R.~T. 1968, {Principles of Stellar Structure: Physical
  principles}, Principles of Stellar Structure No. Bd. 1 (Gordon and Breach)

\bibitem[{D'Cruz {et~al.}(1996)D'Cruz, Dorman, Rood, \& O'Connell}]{D'Cruz1996}
D'Cruz, N.~L., Dorman, B., Rood, R.~T., \& O'Connell, R.~W. 1996, \apj, 466,
  359

\bibitem[{Dorman \& Rood(1993)}]{Dorman1993a}
Dorman, B., \& Rood, R.~T. 1993, \apj, 409, 387

\bibitem[{Dorman {et~al.}(1993)Dorman, Rood, \& O'Connell}]{Dorman1993}
Dorman, B., Rood, R.~T., \& O'Connell, R.~W. 1993, \apj, 419, 596

\bibitem[{Eggleton(1972)}]{Eggleton1972}
Eggleton, P. 1972, \mnras, 156, 361

\bibitem[{Fabrizio \& Bragaglia(2005)}]{Fabrizio2005}
Fabrizio, L.~D., \& Bragaglia, A. 2005, \mnras, 359, 966

\bibitem[{Fontaine \& Chayer(1997)}]{Fontaine1997}
Fontaine, G., \& Chayer, P.\ 1997, Third Conference on Faint Blue Stars,(L. Davis Press, 1997), 169

\bibitem[{Fontaine {et~al.}(2003)Fontaine, Brassard, Charpinet, Green, Chayer,
Bill\`eres, Randall}]{Fontaine2003}
Fontaine, G., Brassard, P., Charpinet, S., {et~al.}\ 2003, \apj, 597, 518

\bibitem[{Fontaine {et~al.}(2006{\natexlab{a}})Fontaine, Brassard, Charpinet, 
  \& Chayer}]{Fontaine2006a}
Fontaine, G., Brassard, P., Charpinet, S., \& Chayer, P.\ 2006{\natexlab{a}}, \memsai, 77, 49 

\bibitem[{Fontaine {et~al.}(2006{\natexlab{b}})Fontaine, Green, Chayer, 
  Brassard, Charpinet, \& Randall}]{Fontaine2006b} 
Fontaine, G., Green, E.~M., Chayer, P., {et~al.} 2006{\natexlab{b}}, Baltic Astronomy, 15, 211 

\bibitem[{Fontaine {et~al.}(2012)Fontaine, Brassard, Charpinet, Green, Randall,
  \& {Van Grootel}}]{Fontaine2012}
Fontaine, G., Brassard, P., Charpinet, S., {et~al.} 2012, \aap, 539, A12

\bibitem[Freytag, Ludwig, \& Steffan(1996)]{fls96} Freytag, B.,
Ludwig, H.-G., \& Steffan, M., 1996, \aap, 313, 497

\bibitem[{Gozzoli \& Tosi(1996)}]{Gozzoli1996}
Gozzoli, E., \& Tosi, M. 1996, \mnras, 283, 66

\bibitem[{Green {et~al.}(2003)Green, Fontaine, Reed, Callerame, Seitenzahl,
  White, Hyde, \O~stensen, Cordes, Brassard, Falter, Jeffery, Dreizler, Schuh,
  Giovanni, Edelmann, Rigby, \& Bronowska}]{Green2003}
Green, E.~M., Fontaine, G., Reed, M.~D., {et~al.} 2003, \apj, 583, L31

\bibitem[{Green {et~al.}(2008)Green, Fontaine, Hyde, For, \& Chayer}]{Green2008}
Green, E.~M., Fontaine, G., Hyde, E.~A., For, B.-Q., \& Chayer, P.\ 2008, 

Hot Subdwarf Stars and Related Objects, 392, 75 

\bibitem[{Green {et~al.}(2011)Green, Guvenen, O'Malley, O'Connell, Baringer,
  Villareal, Carleton, Fontaine, Brassard, \& Charpinet}]{Green2011}
Green, E.~M., Guvenen, B., O'Malley, C.~J., {et~al.} 2011, \apj, 734, 59

\bibitem[{Han {et~al.}(2003)Han, Podsiadlowski, Maxted, \& Marsh}]{Han2003}
Han, Z., Podsiadlowski, P., Maxted, P. F.~L., \& Marsh, T.~R. 2003, \mnras,
  341, 669

\bibitem[{Han {et~al.}(2002)Han, Podsiadlowski, Maxted, Marsh, \&
  Ivanova}]{Han2002}
Han, Z., Podsiadlowski, P., Maxted, P. F.~L., Marsh, T.~R., \& Ivanova, N.
  2002, \mnras, 336, 449

\bibitem[{Heber(1986)}]{Heber1986}
Heber, U. 1986, \aap, 155, 33

\bibitem[{Heber {et~al.}(2000)Heber, Reid, \& Werner}]{Heber2000}
Heber, U., Reid, I., \& Werner, K. 2000, \aap, 363, 198

\bibitem[{Heiter {et~al.}(2014)Heiter, Soubiran, Netopil, \&
  Paunzen}]{Heiter2014}
Heiter, U., Soubiran, C., Netopil, M., \& Paunzen, E. 2014, \aap, 561, A93

\bibitem[{Herwig(2000)}]{Herwig2000}
Herwig, F. 2000, \aap, 360, 952

\bibitem[Herwig, {et~al.}(2014)]{herwig14} Herwig, F., Woodward, P. R., Lin, P. H., Knox, M., \& Fryer, C., 2014, \apj, 792, 3

\bibitem[{Hoyle(1954)}]{Hoyle1954}
Hoyle, F. 1954, \apjs

\bibitem[{Hu {et~al.}(2008)Hu, Dupret, Aerts, Nelemans, Kawaler, Miglio,
  Montalban, \& Scuflaire}]{Hu2008}
Hu, H., Dupret, M., Aerts, C., {et~al.} 2008, \aap, 490, 243

\bibitem[{Hu {et~al.}(2010)Hu, Glebbeek, Thoul, Dupret, Stancliffe, Nelemans,
  \& Aerts}]{Hu2010}
Hu, H., Glebbeek, E., Thoul, A.~A., {et~al.} 2010, \aap, 511, A87

\bibitem[{Hu {et~al.}(2009)Hu, Nelemans, Aerts, \& Dupret}]{Hu2009}
Hu, H., Nelemans, G., Aerts, C., \& Dupret, M.-A. 2009, \aap, 508, 869

\bibitem[{Hu {et~al.}(2011)Hu, Tout, Glebbeek, \& Dupret}]{Hu2011}
Hu, H., Tout, C.~A., Glebbeek, E., \& Dupret, M.-A. 2011, \mnras, 418, 195

\bibitem[{Iglesias \& Rogers(1993)}]{Iglesias1993}
Iglesias, C., \& Rogers, F. 1993, \apj, 412, 752

\bibitem[{Iglesias \& Rogers(1996)}]{Iglesias1996}
---. 1996, \apj, 464, 943

\bibitem[{Iliadis(2007)}]{Iliadis2007}
Iliadis, C. 2007, {Nuclear Physics of Stars} (Wiley-VCH Verlag, 2007)

\bibitem[{Jeffries {et~al.}(2013)Jeffries, Sandquist, Mathieu, Geller, Orosz,
  Milliman, Brewer, Platais, Brogaard, Grundahl, Frandsen, Dotter, \&
  Stello}]{Jeffries2013}
Jeffries, M.~W., Sandquist, E.~L., Mathieu, R.~D., {et~al.} 2013, \aj, 146, 58

\bibitem[{Kaluzny(1998)}]{Kaluzny1998}
Kaluzny, J. 1998, \aaps, 133, 25

\bibitem[{{Kaluzny} {et~al.}(2014){Kaluzny}, {Rozyczka}, {Pych}, \&
  {Thompson}}]{Kaluzny2014}
{Kaluzny}, J., {Rozyczka}, M., {Pych}, W., \& {Thompson}, I.~B. 2014, Acta
  Astronomica, 64, 77

\bibitem[{Kaluzny {et~al.}(2006)Kaluzny, Thompson, Krzeminski, \&
  Schwarzenberg-Czerny}]{Kaluzny2006}
Kaluzny, J., Thompson, I.~B., Krzeminski, W., \& Schwarzenberg-Czerny, a. 2006,
  \mnras, 365, 548

\bibitem[{Kilkenny {et~al.}(1997)Kilkenny, Koen, O'Donoghue, \&
  Stobie}]{Kilkenny1997}
Kilkenny, D., Koen, C., O'Donoghue, D., \& Stobie, R.~S. 1997, \mnras, 285, 640

\bibitem[{Landau \& Lifshitz(1987)}]{LL-FluidMechanics}
Landau, L.~D., \& Lifshitz, E.~M. 1987, {Fluid Mechanics}, 2nd edn.
  (Butterworth-Heinemann, 1987)

\bibitem[{Langer {et~al.}(1983)Langer, Sugimoto, \& Fricke}]{Langer1983}
Langer, N., Sugimoto, D., \& Fricke, K.~J. 1983, \aap, 126, 207

\bibitem[{Meakin \& Arnett(2007)}]{Meakin2007} Meakin, C.~A., \& Arnett, D. 2007, \apj, 667, 448

\bibitem[{Meibom {et~al.}(2009)Meibom, Grundahl, Clausen, Mathieu, Frandsen,
  Pigulski, Narwid, Steslicki, \& Lefever}]{Meibom2009}
Meibom, S.~R., Grundahl, F., Clausen, J.~V., {et~al.} 2009, \aj, 137, 5086

\bibitem[{Mengel {et~al.}(1976)Mengel, Norris, \& Gross}]{Mengel1976}
Mengel, J.~G., Norris, J., \& Gross, P.~G. 1976, \apj, 204, 488

\bibitem[Michaud(1970)]{michaud70} Michaud, G., 1970, \apj, 160, 641

\bibitem[{Michaud {et~al.}(1985)Michaud, Bergeron, Wesemael, \& Fontaine}]{Michaud1985} 
Michaud, G., Bergeron, P., Wesemael, F., \& Fontaine, G.\ 1985, \apj, 299, 741 

\bibitem[Michaud(1991)]{michaud91} Michaud, G., 1991, Ann. Phys., 16, 481

\bibitem[Michaud, Richer, \& Richard(2005)]{mrr05} Michaud, G., Richer, J., \& Richard, O.,2005, \apj, 623, 442

\bibitem[Michaud, Richer, \& Richard(2007)]{mrr07} Michaud, G.,  Richer, J., \& Richard, O.,2007, \apj, 670, 1178

\bibitem[Michaud, Richer, \& Richard(2008)]{mrr08} Michaud, G., Richer, J., \& Richard, O., 2008, \apj, 675, 1223

\bibitem[Michaud, Richer, \& Richard(2011)]{mrr11} Michaud, G.,  Richer, J., \& Richard, O.,2011, \aap, 510, 60


\bibitem[Miller Bertolami {et~al.}(2008) Miller Bertolami, Althaus, Unglaub \& Weiss]{MillerBertolami2008} Miller Bertolami, M.~M., Althaus, L.~G., Unglaub, K. {et~al.} 2008, \aap, 491, 253


\bibitem[Mo\'{c}ak et al.(2009)]{Mocak2009} Mo\'{c}ak, M., M\"{u}ller, E., Weiss, A., \& Kifonidis,K.,2009, \aap, 501, 659

\bibitem[Mo\'{c}ak et al.(2008)]{Mocak2008} Mo\'{c}ak, M., M\"{u}ller, E., Weiss, A., \& Kifonidis,K.,2008, \aap, 490, 265

\bibitem[{Montgomery(1993)}]{Montgomery1993}
Montgomery, K. 1993, \aj, 106

\bibitem[{{\O}stensen {et~al.}(2012){\O}stensen, Degroote, Telting, Vos, Aerts,
  Jeffery, Green, Reed, \& Heber}]{Ostensen2012}
{\O}stensen, R.~H., Degroote, P., Telting, J.~H., {et~al.} 2012, \apj, 753, L17

\bibitem[{Pablo {et~al.}(2012)Pablo, Kawaler, Reed, Bloemen, Charpinet, Hu,
  Telting, \O~stensen, Baran, Green, Hermes, Barclay, O’Toole, Mullally,
  Kurtz, Christensen-Dalsgaard, Caldwell, Christiansen, \&
  Kinemuchi}]{Pablo2012}
Pablo, H., Kawaler, S.~D., Reed, M.~D., {et~al.} 2012, \mnras, 422, 1343

\bibitem[{Paxton {et~al.}(2011)Paxton, Bildsten, Dotter, Herwig, Lesaffre, \&
  Timmes}]{Paxton2011}
Paxton, B., Bildsten, L., Dotter, A., {et~al.} 2011, \apjs, 192, 35

\bibitem[{Paxton {et~al.}(2013)Paxton, Cantiello, Arras, Bildsten, Brown,
  Dotter, Mankovich, Montgomery, Stello, Timmes, \& Townsend}]{Paxton2013}
Paxton, B., Cantiello, M., Arras, P., {et~al.} 2013, \apjs, 208, 4

\bibitem[{Randall {et~al.}(2007)Randall, Green, {Van Grootel}, Fontaine, \&
  Charpinet}]{Randall2007}
Randall, S.~K., Green, E.~M., {Van Grootel}, V., Fontaine, G., \& Charpinet, S.
  2007, \aap, 476, 1317

\bibitem[{Reimers(1975)}]{Reimers1975}
Reimers, D. 1975, Memoires of the Societe Royale des Sciences de Liege, VIII,
  369

\bibitem[{Saffer(1991)}]{1991PhDT........10S}
Saffer, R.~A. 1991, PhD thesis, University of Arizona, Tucson.

\bibitem[{Saffer {et~al.}(1994)Saffer, Bergeron, Koester, \&
  Liebert}]{Saffer1994}
Saffer, R.~A., Bergeron, P., Koester, D., \& Liebert, J. 1994, \apj, 432, 351

\bibitem[{Sandquist {et~al.}(2013)Sandquist, Shetrone, Serio, \&
  Orosz}]{Sandquist2013}
Sandquist, E.~L., Shetrone, M., Serio, A.~W., \& Orosz, J. 2013, \aj, 146, 40

\bibitem[{Saslaw \& Schwarzschild(1965)}]{Saslaw1965}
Saslaw, W., \& Schwarzschild, M. 1965, \apj, 142, 1468

\bibitem[{Schaller {et~al.}(1992)Schaller, Schaerer, Meynet, \&
  Maeder}]{Schaller1992}
Schaller, G., Schaerer, D., Meynet, G., \& Maeder, A. 1992, \aaps, 96, 269

\bibitem[{Schuh {et~al.}(2006)Schuh, Huber, Dreizler, Heber, Toole, Green, \&
  Fontaine}]{Schuh2006}
Schuh, S., Huber, J., Dreizler, S., {et~al.} 2006, \aap, 445, 31

\bibitem[{{Sweigart}(1987)}]{Sweigart1987}
{Sweigart}, A.~V. 1987, \apjs, 65, 95

\bibitem[{Tosi {et~al.}(2007)Tosi, Bragaglia, \& Cignoni}]{Tosi2007}
Tosi, M., Bragaglia, A., \& Cignoni, M. 2007, \mnras, 378, 730

\bibitem[{Tosi {et~al.}(2004)Tosi, {Di Fabrizio}, Bragaglia, Carusillo, \&
  Marconi}]{Tosi2004}
Tosi, M., {Di Fabrizio}, L., Bragaglia, A., Carusillo, P.~A., \& Marconi, G.
  2004, \mnras, 354, 225

\bibitem[{{Van Grootel} \& Charpinet(2008)}]{VanGrootel2008}
{Van Grootel}, V., \& Charpinet, S. 2008, \aap, 488, 685

\bibitem[{{Van Grootel} {et~al.}(2010{\natexlab{a}}){Van Grootel}, Charpinet,
  Fontaine, Green, \& Brassard}]{VanGrootel2010a}
{Van Grootel}, V., Charpinet, S., Fontaine, G., Green, E.~M., \& Brassard, P.
  2010{\natexlab{a}}, \aap, 524, A63

\bibitem[{{Van Grootel} {et~al.}(2010{\natexlab{b}}){Van Grootel}, Charpinet,
  Fontaine, Brassard, Green, Randall, Silvotti, \O~stensen, Kjeldsen,
  Christensen-Dalsgaard, Borucki, \& Koch}]{VanGrootel2010}
{Van Grootel}, V., Charpinet, S., Fontaine, G., {et~al.} 2010{\natexlab{b}},
  \apj, 718, L97

\bibitem[{{Van Grootel} {et~al.}(2013)}]{VanGrootel2013}
{Van Grootel}, V., Charpinet, S., Brassard, P., Fontaine, G., \& 
Green, E.~M.\ 2013, \aap, 553, AA97 

\bibitem[{Viallet {et~al.}(2013)Viallet, Meakin, Arnett, \&
  Moc\'{a}k}]{Viallet2013}
Viallet, M., Meakin, C., Arnett, D., \& Moc\'{a}k, M. 2013, \apj, 769, 1

\bibitem[{Zloczewski {et~al.}(2007)Zloczewski, Kaluzny, Krzeminski, Olech, \&
  Thompson}]{Zloczewski2007}
Zloczewski, K., Kaluzny, J., Krzeminski, W., Olech, A., \& Thompson, I.~B.
  2007, \mnras, 380, 1191


\end{thebibliography}
\bibliographystyle{apj}

\end{document}